\documentclass[acmsmall]{acmart}

\usepackage{graphicx}
\usepackage{amsmath}
\usepackage{amsthm}
\usepackage{booktabs}
\usepackage{algorithm}
\usepackage{algorithmic}
\usepackage{amsfonts}
\usepackage{multirow}
\usepackage{threeparttable}
\usepackage{array}
\usepackage{float}
\usepackage{enumitem}
\usepackage{subfigure}
\usepackage{color}

\AtBeginDocument{%
  \providecommand\BibTeX{{%
    \normalfont B\kern-0.5em{\scshape i\kern-0.25em b}\kern-0.8em\TeX}}}

\begin{document}

\title{A Comprehensive Survey on Automated Machine Learning for Recommendations}

\author{Bo Chen}
\authornote{Both authors contributed equally to this research.}
\affiliation{%
  \institution{Huawei Noah's Ark Lab}
  \country{China}}
\email{chenbo116@huawei.com}

\author{Xiangyu Zhao}
\authornotemark[1]
\affiliation{%
  \institution{City University of Hong Kong}
  \country{HK}}
\email{xianzhao@cityu.edu.hk}

\author{Yejing Wang}
\affiliation{%
  \institution{City University of Hong Kong}
  \country{HK}}
\email{adave631@gmail.com}

\author{Wenqi Fan}
\affiliation{%
  \institution{The Hong Kong Polytechnic University}
  \country{HK}}
\email{wenqifan03@gmail.com}

\author{Huifeng Guo}
\affiliation{%
  \institution{Huawei Noah's Ark Lab}
  \country{China}}
\email{huifeng.guo@huawei.com}

\author{Ruiming Tang}
\affiliation{%
  \institution{Huawei Noah's Ark Lab}
  \country{China}}
\email{tangruiming@huawei.com}

\newcommand{\zxy}[1]{{\color{blue} [xiangyu: #1]}}
\newcommand{\cb}[1]{{\bf \color{cyan} [[cb says ``#1'']]}}
\renewcommand{\shortauthors}{Bo Chen and Xiangyu Zhao, et al.}

\begin{abstract}
Deep recommender systems (DRS) are critical for current commercial online service providers, which address the issue of information overload by recommending items that are tailored to the user's interests and preferences.
They have unprecedented feature representations effectiveness and the capacity of modeling the non-linear relationships between users and items. 
Despite their advancements, DRS models, like other deep learning models, employ sophisticated neural network architectures and other vital components that are typically designed and tuned by human experts. 
This article will give a comprehensive summary of automated machine learning (AutoML) for developing DRS models. 
We first provide an overview of AutoML for DRS models and the related techniques.
Then we discuss the state-of-the-art AutoML approaches that automate the feature selection, feature embeddings, feature interactions, and model training in DRS. We point out that the existing AutoML-based recommender systems are developing to a \textbf{multi-component} joint search with \textbf{abstract} search space and \textbf{efficient} search algorithm.
Finally, we discuss appealing research directions and summarize the survey.
\end{abstract}




\maketitle

\section{Introduction}
\label{sec:intro}
Recent years have witnessed the explosive growth of online service providers~\cite{ricci2011introduction}, including a range of scenarios like movies, music, news, short videos, e-commerces, \textit{etc}~\cite{guo2017deepfm,dien}. This leads to the increasingly serious information overload issue, overwhelming web users.
Recommender systems are effective mechanisms that mitigate the above issue by intelligently retrieving and suggesting personalized items, \textit{e.g.}, contents and products, to users in their information-seeking endeavors, so as to match their interests and requirements better.
With the development and prevalence of deep learning, deep recommender systems (DRS) have piqued interests from both academia and industrial communities
~\cite{zhang2019deep,nguyen2017personalized}, due to their superior capacity of learning feature representations and modeling non-linear interactions between users and items~\cite{zhang2019deep}. %resnick1997recommender  wu2016personal

Most of the existing DRS models feed manually selected features into deep architecture with ``Feature Embedding \& Feature Interaction'' paradigm for recommendation, shown in Figure~\ref{fig:drs}, which consists of several core layers:
1) Input feature layer that feeds the original features into the models; 
2) Feature embedding layer that converts the sparse features into dense representations~\cite{he2014practical};
3) Feature interaction layer that captures the explicit and implicit interactive signals~\cite{guo2017deepfm,pnn,wang2017deep};
4) Output layer that generates the predicted scores for model optimization~\cite{bpr}. 
To construct DRS architectures, the most common practice is to design and tune the different components in a hand-crafted fashion. However, manual development is fraught with these inherent challenges.
\begin{itemize}[leftmargin=*]
\item Manual development requires extensive expertise in deep learning and recommender systems, hindering the development of recommendation.
\item Substantial engineering labor and time cost are required to design task-specific components for various recommendation scenarios. Therefore, plenty of works focus on designing beneficial interaction methods for various scenarios, such as inner product in DeepFM~\cite{guo2017deepfm} and PNN~\cite{pnn}, outer product in CFM~\cite{cfm}, cross operation in DCN~\cite{wang2017deep} and xDeepFM~\cite{lian2018xdeepfm}, and \textit{etc}.
\item Human bias and error can result in sub-optimal DRS components, further reducing recommendation performance. Besides, these manually-designed DRS components have poor generalization, making it difficult to achieve consistent performance on different scenarios.
\end{itemize}

% First, this requires extensive expertise in deep learning and recommender systems. Second, substantial engineering labor and time cost are required to design task-specific components for various recommendation scenarios. Third, human bias and error can result in suboptimal DRS components, further reducing recommendation performance.

Recently, powered by the advances of both theories and technologies in automated machine learning (AutoML), tremendous interests are emerging for automating the components of DRS. Most specifically,  AutoML has been successfully involved into the process of  automatically designing deep recommender systems such as \emph{feature selection, feature embedding search, feature interaction search}, and \emph{model training}, as illustrated in Figure~\ref{fig:auto-rec}. 
By involving AutoML for the deep recommender systems, different models can be automatically designed according to various data, thus improving the prediction performance and enhancing generalization.
%The initial motivation of these work is to enable the amateurs can obtain comparable performance of DRS models through AutoML tools rather than becoming an experts in the domain of DRS. 
Besides, it is helpful to eliminate the negative influence for DRS from human bias and error, as well as reduce artificial and temporal costs significantly.
% AutoML enables non-experts to create DRS models without becoming domain experts, eliminates the possibility of human bias and error, and reduces artificial and temporal costs significantly. 
Typically, these works can be divided into the following categories according to the different components in DRS:

% DRS frameworks are composed of the following critical components: 

\begin{figure}[!t]
    \centering
    \setlength{\belowcaptionskip}{-0.1cm}
    \hspace*{-6mm}\includegraphics[width=0.95\textwidth]{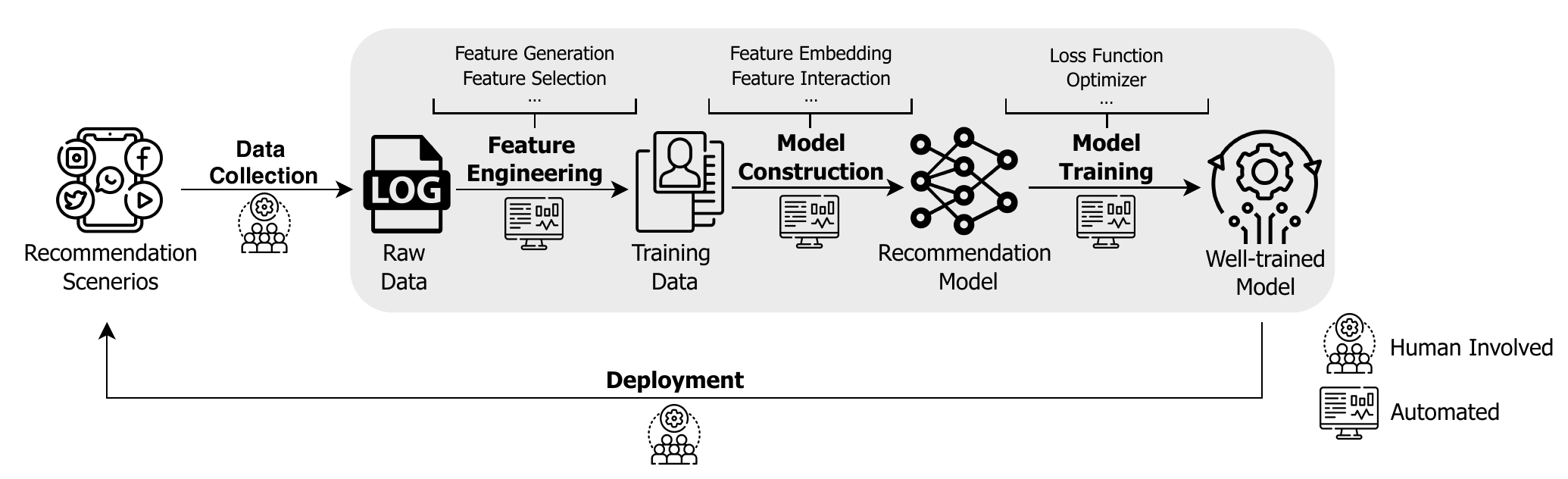}
    \caption{AutoML for Building Deep Recommender Systems. }
 \vspace{-0.3cm}
    \label{fig:auto-rec}
\end{figure}

\begin{figure}[!t]
    \centering
    \setlength{\belowcaptionskip}{-0.1cm}
    \hspace*{-6mm}\includegraphics[width=0.95\textwidth]{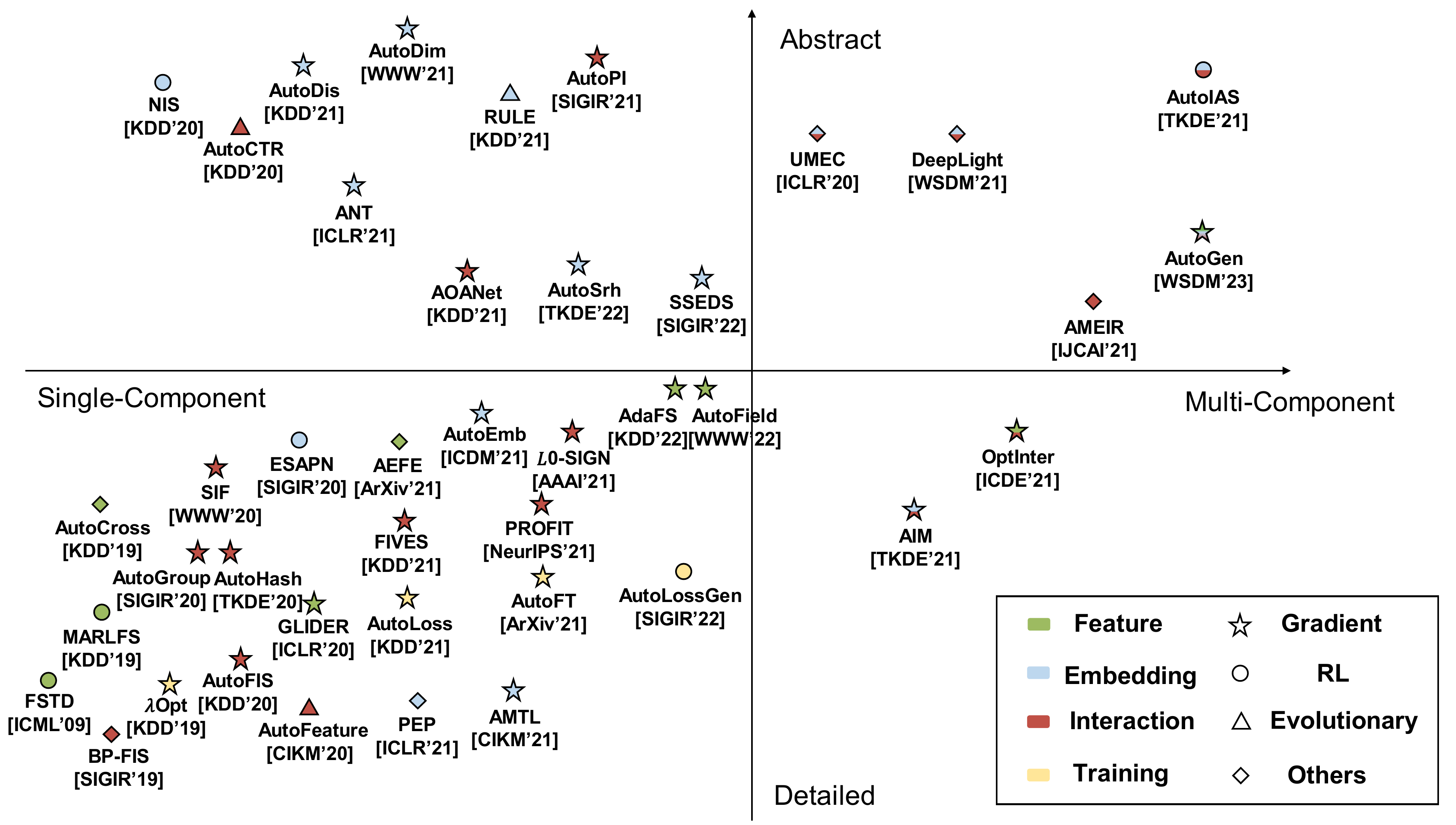}
     \caption{The trend of AutoML for recommender system. The search space is divided from two orthogonal dimensions, \textit{i.e.}, single-component or multi-component search space, and abstract or detailed search space.
     Colors represent the components of DRS and shapes denote different search strategy.}
 \vspace{-0.3cm}
    \label{fig:trend}
\end{figure}

\begin{itemize}[leftmargin=*]
\item \textbf{Feature Selection}: This is the process of selecting a subset of the most predictive and relevant features (or generated features) for subsequent DRS models. 
By eliminating the redundant or irrelevant features, feature selection can help enhance the recommendation performance and accelerate DRS model training~\cite{nadler2005prediction}.

\item\textbf{Feature Embedding}: 
Feature embedding layer is utilized to transform the high-dimensional and extremely sparse raw features into dense representations.
AutoML technique is utilized to dynamically search the optimal embedding sizes for improving prediction accuracy, saving storage space, and reducing model capacity. 
% Typically, the features for DRS are high-dimensional and extremely sparse. 
% Most recommendation models first transform the raw features into one-hot vectors and then embed them as dense representations via the feature embedding layer.
% AutoML technique is utilized to dynamically search the optimal embedding sizes for improving prediction accuracy, saving storage space, and reducing model capacity. 
%Embedding parameters usually dominate DRS model parameters, which naturally play an essential role in recommendation performance.

\item\textbf{Feature Interaction}: 
%The prediction in DRS models cannot be expressed as the sum of the effects of all input features separately, because the interaction of two features would alter their individual feature effects. For example,  users often download food delivery apps at meal time, in which  interactions between app category and time-stamp is a highly predictive feature. Therefore, mining predictive feature interactions is one of the key techniques to boost recommendation quality.
Effectively modeling predictive feature interactions is critical for boosting the recommendation quality of DRS. % because the interaction of two features would alter their individual feature effects. %For example, users often download food delivery apps at mealtime, in which interactions between app category and time-stamp is a highly predictive signal. 
Therefore, some AutoML-based works are devoted to exploring beneficial feature interactions with proper interaction functions.

% \item \textbf{Hidden Layer}: This DRS component aims to conduct nonlinear transformations and learn the higher-order representations of input feature embeddings. Besides, the last layer of this component (e.g, output layer) will produce the prediction for the specific downstream recommendation tasks. 
% Different from the deep neural networks in CV and NLP tasks, DRS models have relatively shallower hidden layers with simpler architectures.

\item \textbf{Model Training}: 
In addition to the above components of DRS models, model training also has a crucial impact on DRS performances, including data pipeline, model optimization and evaluation, hardware infrastructure as well as deployment, .

% \item \textbf{System Design}: 
% In addition to the above components of DRS models, system design also has a crucial impact on DRS performances, including hardware infrastructure, data pipeline, and information transfer, as well as implementation, deployment, optimization, and evaluation. 
\end{itemize}

% To construct DRS architectures, the most common practice is to design and tune the aforementioned components in a hand-crafted fashion. However,   manual development is fraught with three inherent challenges. First, this requires extensive expertise in deep learning and recommender systems. Second, substantial engineering labor and time cost are required to design task-specific components for various recommendation scenarios. Third, human bias and error can result in suboptimal DRS components, further reducing recommendation performance.

% \red{Recently, powered by the advances of both theories and technologies in automated machine learning (AutoML), tremendous interest are emerging for automating the components of DRS. The initial motivation of these work is to enable the amateurs can obtain comparable performance of DRS models through AutoML tools rather than becoming an experts in the domain of DRS. 
% Besides, it is helpful to eliminate the negative influence for DRS from human bias and error and reduce artificial and temporal costs significantly.
% % AutoML enables non-experts to create DRS models without becoming domain experts, eliminates the possibility of human bias and error, and reduces artificial and temporal costs significantly. 
% As shown in Tabel~\ref{tab:comp}, we summarize these research from the perspective of search space and search strategy, which are two critical factors for the AutoML RS approaches.\wq{not smooth}}

We summarize these researches from the perspective of search space and search strategy, which is depicted in Figure~\ref{fig:trend}. Considering whether the search space is single-component or multi-component, the majority of work focus on searching single component in DRS automatically, such as feature selection~\cite{wang2022autofield,glider}, embedding dimension search~\cite{autoemb,nis}, \textit{etc}, while leaving other components as fixed manual designs. Recently, several studies are devoted to search multiple components comprehensively~\cite{aim,autoias}, thus reducing the involvement of expert experience in DRS design. 
Another divided dimension is the abstraction degree of the depicted search space. Most work design a detailed search space for each component, such as searching fine-grained embedding dimension for each feature~\cite{pep,amtl}, exploring explicit feature interactions~\cite{autofis,autofeature}, \textit{etc}. To improve search efficiency, recently some studies reduce the detailed search space into an abstract one, such as searching a unified embedding dimension for a group of features~\cite{sseds,autosrh}, searching high-order feature interactions over the whole feature sets implicitly~\cite{autoctr,autopi}.
Besides, we also present their search algorithm, including gradient-based, RL-based, evolutionary-based, and others.
From Figure~\ref{fig:trend}, we observe the following development directions of existing AutoML-based recommendation models:
\begin{itemize}[leftmargin=*]
\item The existing AutoML-based work evolves from single-component search to \textbf{multi-component} joint search. 
\item The search space of these AutoML-based work develops from detailed to \textbf{abstract} for shrinking search space and improving search efficiency.
\item The search algorithm of existing work is mainly based on gradient-based methods~\cite{liu2018darts}, thus providing \textbf{efficient} model searching and training mode.
\end{itemize}

This survey is to provide a literature overview on the advances of AutoML for constructing DRS architectures.
To be specific, we first provide an overview of AutoML techniques and present the architecture of deep recommender systems. %DRS models and
Then, we discuss the state-of-the-art AutoML approaches that automate the feature selection, feature embeddings, feature interactions, and model training in DRS models, as well as introduce some jointly-designed works. 
Finally, we discuss some emerging topics and the appealing directions that can bring this research field into a new frontier.

\newcommand{\w}[1]{{\color{red}#1}}

\section{AutoML Preliminaries}
In this section, we will first give an overview of AutoML techniques and introduce some representative methodologies. Then, we discuss existing surveys on AutoML techniques and point out our distinguished features.
\subsection{Technique Overview}
With the given problem descriptions and datasets, the goal of Automated Machine Learning (AutoML) techniques is to automatically construct machine learning solutions for time-consuming and iterative real-world tasks.
It has shifted the model design mechanism from hand-crafted to automatic, enabling unparalleled prospects for deep learning model construction. AutoML frameworks are typically comprised of three following components:
\begin{itemize}[leftmargin=*]
\item \textbf{Search Space}. The search space defines a group of candidate operations and their connections that enable appropriate model designs to be formed. For DRS, different components contain diverse search spaces involved with human prior knowledge, such as input features, embedding dimensions, feature interactions, and network operations. %Typically, human prior knowledge is incorporated to reduce the search space, unavoidably, with human biases.
\item \textbf{Search Strategy}. The search strategy specifics how to conduct the efficient exploration of the search space and find out the optimal architectures, which typically contains reinforcement learning (RL)~\cite{kaelbling1996reinforcement},  gradient-based optimization~\cite{ruder2016overview}, evolutionary algorithms~\cite{qin2008differential}, Bayesian optimization~\cite{snoek2012practical}, random search~\cite{bergstra2012random},  \textit{etc}. 
% or grid hsu2003practical
%In our survey, we categorize search methods into following types: (i) gradient: gradient-based methods like DARTS~\cite{liu2018darts}; (ii) regularization: search architectures depending on regularizers; (iii) RL: reinforcement learning methods; (iv) evolutionary: construct networks by evolutionary algorithms (v) Bayesian: optimizing structures by Bayesian optimization; (vi) random search: random NAS as one-shot random search~\cite{li2020random}.} %这部分是根据后面出现过的简称写的
% We briefly introduce some popular methods in this section.  
\begin{itemize}[leftmargin=*]
\item \textbf{Reinforcement Learning (RL)}: %RL 
General RL always constructs the problem with a pair of environment and agent, where the agent takes actions $a_t$ according to its policy $\pi$ and the state $s_t$ of the environment at time $t$. Afterwards, a reward $r_{t+1}$ is assigned to the agent to encourage profitable actions, and the environment state is updated to $s_{t+1}$. The general goal of RL is to maximize the accumulated reward $R$ and learn the optimal policy $\pi^*$ for the agent:
\begin{align}
    R&=\sum_{t=0}^{T-1} \gamma_t r_{t+1} \nonumber \\
    \pi^*&=\underset{\pi}{\operatorname{argmax}}\, \mathbb{E}(R|\pi), \label{eq:policy}
\end{align}
where $\gamma_t$ is the discount factor, and $\mathbb{E}(\cdot)$ stands for the expectation operator. For AutoML, \citet{zoph2017neural} first formulate the neural architecture search (NAS) problem from the perspective of RL. They define the environment as all possible neural architectures and encode distinct structures as numeric numbers, the state of which is static. A recurrent neural network (RNN) is set as the agent, and the policy is its parameters $\theta_c$. For a single interaction, the RNN first samples a child neural architecture from the whole environment by generating a sequence of numbers. Then, the sampled network is trained to convergence, and the corresponding accuracy is obtained as the reward. Finally, $\theta_c$ is updated by policy gradient:
 \begin{align}
     \nabla_{\theta_c} J(\theta_c)&=\sum_{t=1}^T \mathbb{E}_{P(a_{1: T} ; \theta_c)}\left[\nabla_{\theta_c} \log P(a_t| a_{(t-1): 1} ; \theta_c) R\right] \nonumber \\
     &\approx \frac{1}{m} \sum_{k=1}^m \sum_{t=1}^T \nabla_{\theta_c} \log P\left(a_t \mid a_{(t-1): 1} ; \theta_c\right) R_k,
 \end{align}
where $J(\theta_c)$ is the expected reward $\mathbb{E}(R|\theta_c)$ in Equation (\ref{eq:policy}). $P(a_{1: T} ; \theta_c)$ is the probability of generating the child network with action $\{a_1,\dots, a_T\}$ under the parameter $\theta_c$ and $P(a_t | a_{(t-1): 1} ; \theta_c)$ is the conditional probability for taking the action $a_t$ given all previous actions. The approximate result is the average gradient of $m$ interactions. 
This strategy requires a large amount of computation. Thus, subsequent efforts focus on designing efficient NAS methods with parameter sharing strategy~\cite{pham2018efficient} or progressive architecture generation~\cite{liu2018progressive}.

\item \textbf{Gradient-based Optimization}: % Gradient 
The search space of AutoML works is usually discrete, which prevents researchers from applying the efficient gradient descent algorithm. \citet{liu2018darts} innovatively relax the discrete search space to a continuous one. To be specific, suppose that the input is $x$ and all operation candidates for position $\{i,j\}$ are $\{o^1(\cdot),\dots,o^k(\cdot),\dots,o^K(\cdot)\}$, previous works only select one of them for  model construction during the search stage, and pick the best-performed architecture as the search result. However, DARTS~\cite{liu2018darts} returns a moderate result in searching:
\begin{equation}
\overline{o}_{i,j}(x)=\sum_{k=1}^{K} \frac{\exp \left(\alpha^{k}_{i,j}\right)}{\sum_{l=1}^{K} \exp \left(\alpha^{l}_{i,j}\right)} o^k(x),
\label{eq:darts_softmax}
\end{equation}
where $\alpha^{k}_{i,j}$ is the score for $o^k(\cdot)$. It is noteworthy all operations are involved in Equation (\ref{eq:darts_softmax}). Operations with higher scores $\alpha^{k}_{i,j}$ are regarded as more effective. To construct the final search result, the operation with the largest scores for each index $\{i,j\}$ is selected. This strategy introduces extra architecture parameters $\alpha^{k}_{i,j}$, denoted as $\mathcal{A}$. Together with model parameters $\mathcal{W}$ (the parameter for candidate operations), authors design a bi-level optimization problem and alternatively update two sets of parameters:
\begin{equation}
\begin{aligned}
\label{eq:darts_bilevel}
&\min _{\mathcal{A}} \mathcal{L}_{\text {val}}\left(\mathcal{W^{*}}(\mathcal{A}), \mathcal{A}\right)\\
&\text { s.t. }\mathcal{ W^{*}}(\mathcal{A})=\arg \min _{\mathcal{W}} \mathcal{L}_{\text {train}}\left(\mathcal{W}, \boldsymbol{\mathcal{A}}^{*}\right),
\end{aligned}
\end{equation}
where $\mathcal{ W^{*}}, \boldsymbol{\mathcal{A}}^{*}$ are the optimization result and $\mathcal{L}_{\text {train}},\mathcal{L}_{\text {val}}$ are losses on the training set and validation set, respectively. Plenty of automated recommender systems are constructed based on this solution~\cite{wang2022autofield,yang2021autoft,zhao2021autoloss,autodim,autoemb,lin2022adafs}.
\item \textbf{Evolutionary Algorithms}: Evolution algorithms are inspired by biological evolution in nature. For AutoML, evolutionary methods always encode the search space, \textit{e.g.}, neural architectures, as numeric values (direct encoding) or compact vectors (indirect encoding by auxiliary networks). Then, they iteratively select promising candidates to generate offspring models by interweaving selected parent models (inheriting half genetic information) and randomly mutating the results. After each round of evolution, old or bad-performing models are discarded due to computation resource limitations. The optimal framework is finally obtained with the stopping criteria achieved.
\item \textbf{Bayesian Optimization}: Bayesian optimization is a kind of sequential model-based optimization strategy popular for both hyper-parameter optimization~\cite{falkner2018bohb} and neural architecture search~\cite{thornton2013auto,camero2021bayesian}. Bayesian optimization strategies are generally constructed based on Gaussian Process~\cite{rasmussen2003gaussian}, random forest~\cite{ho1995random}, or tree-structured Parzen estimator~\cite{bergstra2011algorithms}. Bayesian optimization frameworks repeatedly train these models to predict the evaluation results of generated neural architectures or selected hyper-parameters and finally efficiently obtain the promising neural frameworks with an accurate model.
\item \textbf{Random Search}:
With the emergence of complex AutoML methods, random search was once considered an inaccurate strategy. However, \citet{li2020random} suggests a novel random search strategy with a weight-sharing~\cite{pham2018efficient} and early-stopping policy, which beats all baselines and obtains state-of-the-art performance. Specifically, they divide the whole framework into several nodes as DARTS~\cite{liu2018darts}. For each node, they identify the candidate sets for the input and corresponding operations and uniformly sample operations as the final decision. With sampled architectures, the shared model weights are trained for several epochs. The final output is the best-performed architectures with the converged shared weights selected from previous sampling results. This novel random search is effective and easy to reproduce.
\end{itemize}
%The search strategy specifics how to conduct an efficient exploration of the search space and find out the optimal architectures, which typically contains gradient-based optimization~\cite{ruder2016overview}, reinforcement learning (RL)~\cite{kaelbling1996reinforcement}, evolutionary algorithms~\cite{qin2008differential}, Bayesian optimization~\cite{snoek2012practical}, random search~\cite{bergstra2012random}, \textit{etc}.
\item \textbf{Performance Estimation Strategy}. Estimating the performance of specific candidate architectures sampled from the massive search space is vital for generating effective deep architectures. Various strategies for efficient performance estimation have been proposed to reduce the computational cost of repeatedly training and estimating over these candidates, such as weight sharing and network morphism.
%Performance estimation is the process of estimating the performance of sampled candidate architectures from the massive search space. To reduce the computational cost of training and estimating over these candidates, various strategies for performance estimation have been proposed, such as weight sharing~\cite{pham2018efficient} and network morphism~\cite{elsken2017simple}.
\end{itemize}

% Next, we present some widely-applied strategies:

\subsection{A Road-map for AutoML Surveys} 

In this section, we collect various surveys on AutoML and discuss their differences. Then, we summarize our major contributions.

\begin{table}[]
\caption{AutoML surveys. We use ``NAS'' for neural architecture search, ``HPO'' for hyper-parameter optimization, ``FE'' for feature engineering, ``DP'' for data preparation, ``AS'' for algorithm selection, ``RS'' for recommender systems, ``FS'' for feature selection, ``FES'' for feature embedding search, ``FIS'' for feature interaction search.}
\label{tab:surveys} 
\resizebox{0.95\textwidth}{!}{%
\begin{tabular}{@{}lllll@{}}
\toprule
\textbf{Survey} &
  \textbf{Year} &
  \textbf{Range} &
    \textbf{Taxonomy} &
  \textbf{Framework} \\ \midrule
\cite{jaafra2018review} &
  2018 &
  NAS &
  Technique &
  Meta-modeling,  RL methods for NAS \\
\cite{elsken2019neural} &
  2019 &
  NAS &
  Component &
  Search Space, Search Strategy, Evaluation Strategy \\
\cite{wistuba2019survey} &
  2019 &
  NAS &
  Component &
  Search Space, Search Strategy, Evaluation Strategy \\
\cite{weng2020nas} &
  2020 &
  NAS &
  Component &
  Search Space, Search Strategy, Evaluation Strategy \\
\cite{ren2021comprehensive} &
  2021 &
  NAS &
  Challenge-Solution &
  Search Space, Search Strategy, Training Strategy \\%Architecture Recycling, Incomplete Training
\cite{yu2020hyper} &
  2020 &
  HPO &
  Component &
  Search Space, Search Strategy, Early-stopping Policy \\ \midrule
\cite{yao2018taking} &
  2018 &
  FE, HPO, NAS &
  Procedure &
  Problem Definition, Optimizer, Evaluator \\
\cite{elshawi2019automated} &
  2019 &
  DP, HPO, NAS &
  Technique &
  Meta Learning,  NAS, HPO, Tools \\
\cite{zoller2021benchmark} &
  2021 &
  DP, FE, HPO &
  Problem &
  HPO, Data Cleaning, Feature Engineering \\
\cite{chen2021techniques} &
  2021 &
  FE, HPO, NAS &
  Problem &
  FE, HPO, NAS \\
\cite{he2021automl} &
  2021 &
  DP, FE, HPO, NAS &
  Problem &
  DP, FE, Model Generation, Model Evaluation \\
\cite{dong2021automated} &
  2021 &
  FE, NAS, HPO, Deploy, Maintain &
  Problem &
  FE, NAS, HPO, Deployment, Maintenance \\ \midrule
\cite{afshar2022automated} &
  2022 &
  AS, HPO, Meta-Learning,  NAS &
  Problem &
   AS, HPO, Meta-Learning, NAS \\
\cite{parker2022automated} &
  2022 &
  Task Design, AS, NAS, HPO &
  Problem &
   Task Design, AS, NAS, HPO \\
\cite{zhang2021automated} &
  2021 &
  HPO, NAS &
  Problem &
   HPO, NAS \\
\cite{waring2020automated} &
  2020 &
  FE, HPO, NAS &
  Problem &
  FE, HPO, NAS \\
\cite{mustafa2021automated} &
  2021 &
  DP, FE, AS &
  Application &
  AutoML in Healthcare Industry, AutoML for Clinical Notes \\
\cite{zheng2022automl} &
  2022 &
  FE, HPO, NAS... &
    Problem &
  FES, FIS, Multiple components, Other component \\ \midrule
Ours &
  2023 &
  DP, FE, HPO, NAS... &
  RS Components &
  FS, FES, FIS, Training, Comprehensive \\ \bottomrule
\end{tabular}%
}
\end{table}

The collected surveys are listed in Table~\ref{tab:surveys}. The definition of acronyms such as ``NAS'' is explained in the caption of the table. In addition to survey titles, we also include their publication times (Year), the problems or applications they address (Range), their classification standard (Taxonomy), and their article organization (framework).  
The term ``Range'' covers general AutoML issues like ``NAS'' and ``HPO'', as well as specific procedures as ``Task Design'' for AutoRL~\cite{parker2022automated}. 
For ``Taxonomy'', we conclude classification standard of each survey to identify their connections and differences. 
As for ``Framework'', we summarize organization according to survey structures.

In Table~\ref{tab:surveys}, surveys in the first two blocks focus primarily on the traditional AutoML problem, \textit{i.e.}, AutoML for computer vision tasks. 
Surveys in the first block collect papers in a single domain such as ``NAS'' or ``HPO''. The majority of them~\cite{elsken2019neural,wistuba2019survey,weng2020nas,yu2020hyper} categorize the collected works from the perspective of ``Component'' of AutoML, which includes search space, search strategy, and evaluation strategy. %(`training strategy' for Survey\cite{yu2020hyper}
To be specific, \citet{jaafra2018review} classified existing NAS works as meta-modeling methods and RL methods. \citet{ren2021comprehensive} inventively review NAS works from the view of challenges and corresponding solutions, including modular search space, continuous search strategy, neural architecture recycling, and incomplete training methods. 
The surveys in the second block cover a broader range of AutoML techniques applicable to a greater number of practical problems, such as automated feature engineering.
The earliest work~\cite{yao2018taking} introduces AutoML articles in a problem-solving order. They first define the search space for each problem, and then demonstrate how to search for the optimal solution within the corresponding search space (``Optimizer'') and evaluate the solution (``Evaluator''). %, which is analogous to categorizing AutoML works from the view of AutoML components. 
\citet{elshawi2019automated} review AutoML techniques and provide readers with various methods, including meta-learning, HPO and NAS methods, as well as implemented tools for practice. 
The remaining surveys in the second block~\cite{zoller2021benchmark,he2021automl,chen2021techniques,dong2021automated} review AutoML papers primarily from the perspective of problems to be solved. It is noteworthy that \citet{dong2021automated} noticed the automated deployment and maintenance and suggested ten criteria to assess AutoML works in both individual publications and broader research areas.

Recent years also witnessed several AutoML surveys for specific research areas, including automated reinforcement learning~\cite{afshar2022automated,parker2022automated}, graph neural networks~\cite{zhang2021automated}, healthcare~\cite{waring2020automated,mustafa2021automated}, and recommender systems~\cite{zheng2022automl}. The vast majority of them introduce works by different automation problems~\cite{afshar2022automated,parker2022automated,zhang2019deep,waring2020automated,zheng2022automl}. For example, \citet{mustafa2021automated} review AutoML applications in healthcare, such as industry and clinical notes. 
As a AutoML survey for DRS, compared with the survey~\cite{zheng2022automl}, we categorize AutoML works from the view of DRS components, \textit{i.e.}, feature selection (FS), feature embedding search (FES), feature interaction search (FIS), model training (Training). This contributes a greater breadth of coverage than \citet{zheng2022automl}. Moreover, we have a fine-grained classification for each DRS component, \textit{e.g.}, raw feature selection and generated feature selection for FS, and full embedding search and raw/column-wise embedding search for FES. 
This taxonomy manner enables our survey to include all of AutoML works for DRS. In fact, issues with the survey~\cite{zheng2022automl} are included in our survey. Particularly, embedding dimension search in~\citet{zheng2022automl} is contained in column-wise or full embedding search problem in our survey. Feature interation search and feature interaction function search are contained in our feature interaction section. Loss function search\footnote{Feature interaction function search and loss function search belongs to ``other component'' in survey~\cite{zheng2022automl}.} of \citet{zheng2022automl} is a part of our model training section. We conclude our contribution as follows:
\begin{itemize}[leftmargin=*]
    \item To the best of our knowledge, our survey is the first to conduct comprehensive and systematic review of AutoML technologies for DRS according to various DRS components;
    \item Compared with existing surveys, we propose a fine-grained taxonomy manner in each section of DRS component, and provide insights into each of them;
    \item We analyze multiple emerging topics and unexplored problems separately in an effort to identify promising future directions.
\end{itemize}
\section{Preliminary for Deep Recommender System}
\label{preliminary}
\begin{figure}[!t]
    \centering
    \setlength{\belowcaptionskip}{-0.1cm}
    \hspace*{-6mm}\includegraphics[width=0.55\textwidth]{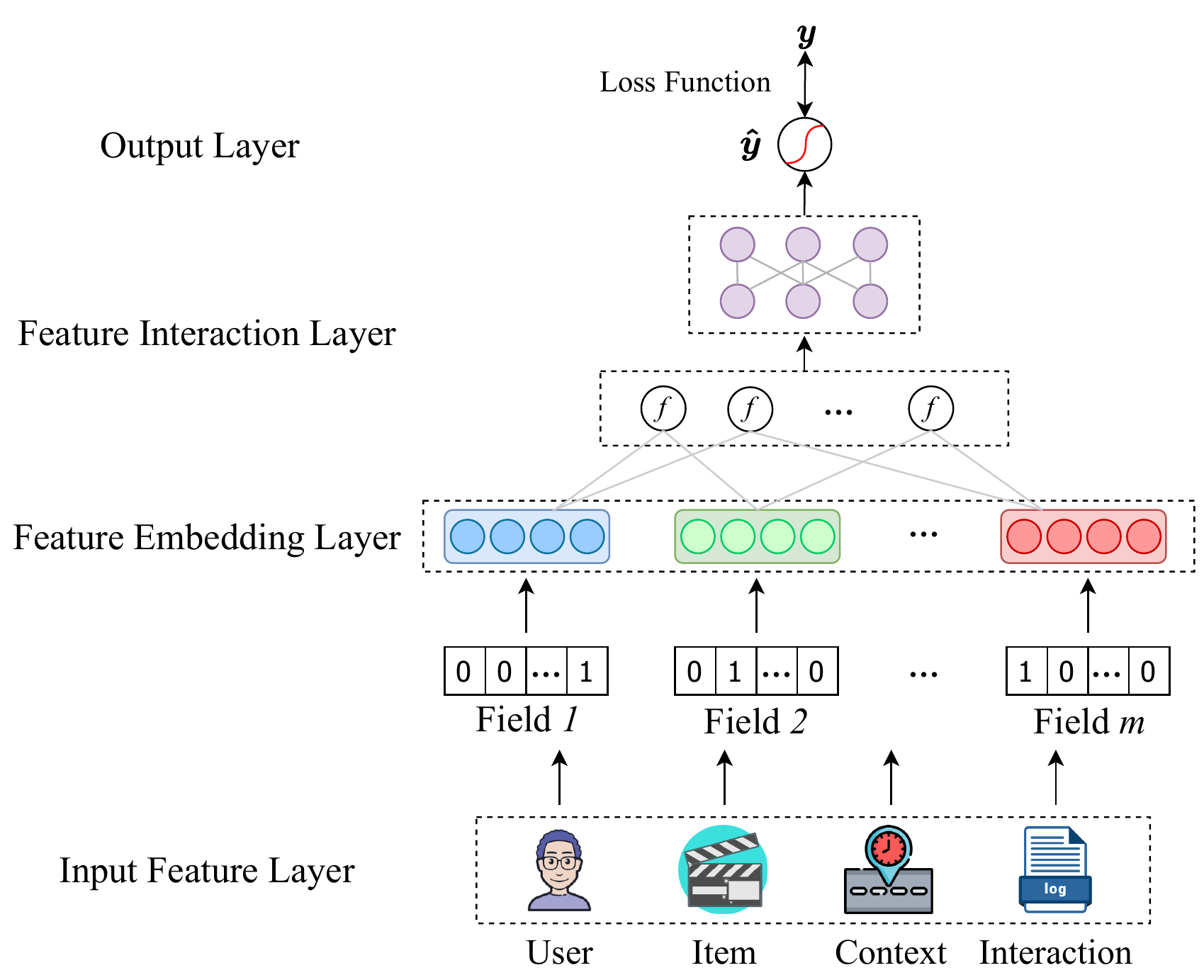}
    \caption{The overview of deep recommender system.}
 \vspace{-0.3cm}
    \label{fig:drs}
\end{figure}

Recently, with the development of deep learning, neural network-based recommender systems become mainstream rapidly. Abstractly, these deep recommendation models follow the ``Feature Embedding \& Feature Interaction'' paradigm, shown in Figure~\ref{fig:drs}, which consists of several core layers~\cite{autodis}: \textit{Input Feature Layer}, \textit{Feature Embedding Layer}, \textit{Feature Interaction Layer}, and \textit{Output Layer}. 

\noindent{\textbf{Input Feature Layer}}

Recommender systems are designed to precisely infer user's preferences over the candidate items based on the historical records. Therefore, the input features for the recommendation models include user profile features (\textit{e.g.}, gender, age), item attribute features (\textit{e.g.}, name, category), contextual features (\textit{e.g.}, weekday, location), as well as their combinatorial features~\cite{autodis,guo2017deepfm}. These features can be divided into enumerable categorical features and infinite numerical features, where numerical features are commonly discretized into categorical features with manually designed rules. 
For clarify, we use ``feature field'' to represent a class of features and ``feature value'' to represent a certain value in a specific feature field. 

Generally, for commercial recommender systems, feature engineers extract and construct plenty of features from the historical records. Based on the pre-processed features, a large amount of manpower is used to manually select informative features as input to the deep recommendation models, which is a time-consuming and labor-intensive process.

\noindent{\textbf{Feature Embedding Layer}}

Assume that $m$ feature fields are selected as inputs, for the $i^{th}$ feature field, each feature value is converted into high-dimensional vector $\mathbf{x}_i \in \mathbb{R}^{V_i}$ via field-wise one-hot encoding~\cite{he2014practical}, where $\mathbf{x}_i$ is the one-hot vector and $V_i$ is the vocabulary size (number of feature values) of the $i^{th}$ field. 
Then, a feature embedding layer, parameterized as $g_{\boldsymbol{\phi}}(\cdot)$, is applied to transform the high-dimensional sparse one-hot vectors into dense latent space for learning feature representations. Specifically, for the $i^{th}$ feature field, a field-wise embedding table $\mathbf{E}_i$ is assigned and the feature embedding vector $\mathbf{e}_i$ can be obtained by embedding look-up:
\begin{equation}
\label{embedding}
    \mathbf{e}_i = \mathbf{E}_i \cdot \mathbf{x}_i,
\end{equation}
where $\mathbf{E}_i \in \mathbb{R}^{V_i\times d}$ is the embedding matrix for the $i^{th}$ field and $d$ is the embedding size. 
Habitually, the embedding dimensions of different fields are set to the same depending on manual experience and the embedding matrices of the $m$ feature fields are concatenated into a global embedding matrix $\mathbf{E} \in \mathbb{R}^{V\times d}$, where $V=\sum_{i=1}^mV_i$ is the total vocabulary size of the $m$ feature fields.
Therefore, the feature embeddings can be presented as:
\begin{equation}
\label{embedding}
    \mathbf{e} = g_{\boldsymbol{\phi}}(\mathbf{x}) = [\mathbf{e}_1, \mathbf{e}_2, \dots, \mathbf{e}_m],
\end{equation}
where $\boldsymbol{\phi}$ is the parameters in the embedding layer. 

\noindent{\textbf{Feature Interaction Layer}}

Based on the learned feature embeddings, feature interaction layers, parameterized as $f_{\boldsymbol{\varphi}}(\cdot)$ where $\boldsymbol{\varphi}$ is the parameters in the feature interaction layer, are deployed to capture informative interaction signals among these features explicitly and implicitly~\cite{guo2017deepfm}. Concretely, the feature interaction layers can be divided into two kinds: 
\begin{itemize}[leftmargin=*]
\item \textbf{Explicit interaction modeling}: Modeling fine-grained feature interactions explicitly with pre-defined interaction function, mainly factorization-based models (\textit{e.g.}, FM~\cite{fm} and PNN~\cite{pnn}). Various interaction functions are deployed to model the feature interactions, such as inner product in DeepFM~\cite{guo2017deepfm} and PNN~\cite{pnn}, outer product in CFM~\cite{cfm}. Taking the second-order feature interaction ($p=2$, where $p$ is the number of order) as an example, for a pair of feature embeddings $\mathbf{e}_i$ and $\mathbf{e}_j$, their interaction can be denoted as $f_{\boldsymbol{\varphi}}(\mathbf{e}_i, \mathbf{e}_j)$. 
% \begin{equation}
% \label{embedding}
%     f_{\boldsymbol{\varphi}}(\mathbf{e}_i, \mathbf{e}_j) = \mathbf{E}_i \cdot \mathbf{x}_i,
% \end{equation}
However, when modeling high-order feature interactions ($p>2$), the generated interactions $\mathcal{C}_m^p$ bring a large computational burden.
\item \textbf{Implicit interaction modeling}: Taking the features as a whole and performing coarse-grained high-order interactions implicitly. To effectively capture high-order interactive signals, plenty of interaction layers are developed in an implicit manner, expressed as $f_{\boldsymbol{\varphi}}(\mathbf{e})$, such as cross layer in DCN~\cite{wang2017deep}, compressed interaction layer in xDeepFM~\cite{lian2018xdeepfm}, and fully connected layer in DNN.
\end{itemize} 

\noindent{\textbf{Output Layer}}

After the feature interaction layer, the predicted score $\hat{y}$ can be obtained by:
\begin{equation}
\label{embedding}
    \hat{y} = f_{\boldsymbol{\varphi}}(g_{\boldsymbol{\phi}}(\mathbf{x})).
\end{equation}
The deep recommendation model is optimized in a supervised manner by minimizing the difference between predicted score $\hat{y}$ and ground truth label $y$ with various loss functions, such as binary cross-entropy (BCE) loss $\mathcal{L}_{BCE}$, bayesian personalized ranking (BPR) loss $\mathcal{L}_{BPR}$~\cite{bpr}, and mean square error (MSE) loss $\mathcal{L}_{MSE}$.
\begin{equation}
\begin{aligned}
\label{bce}
    \mathcal{L}_{BCE} &= -\frac{1}{N}\sum_{i=1}^{N}(y_{i}log(\hat{y}_{i}) + (1-y_{i})log(1-\hat{y}_{i}) ) + \lambda_{L_2} \Vert \boldsymbol{\phi} \Vert_2,\\
    \mathcal{L}_{BPR} &= -\frac{1}{N}\sum_{i=1}^{N}(\frac{1}{N_s}\sum_{j=1}^{N_s}log(\sigma(\hat{y}_{i}-\hat{y}_{j}))) + \lambda_{L_2} \Vert \boldsymbol{\phi} \Vert_2,\\
    \mathcal{L}_{MSE} &= \frac{1}{N}\sum_{i=1}^{N}(y_{i}-\hat{y}_{i})^2 + \lambda_{L_2} \Vert \boldsymbol{\phi} \Vert_2,\\
\end{aligned}
\end{equation}
where $N$ is the number of instances, $N_s$ is the number of negative samples for BPR loss, and $\lambda_{L_2}$ is the hyper-parameter for $L_2$ regularization.

As mentioned above, neural network-based recommender systems involve a lot of manual design experience, hindering the development of recommendations and resulting in sub-optimal performance. 
To overcome these issues, plenty of AutoML-based methods are proposed to automate the recommendation models, ranging from input feature selection (in Section~\ref{sec:feature}), feature embedding search (in Section~\ref{sec:embedding}), feature interaction search (in Section~\ref{sec:fi}), to model training (in Section~\ref{sec:training}). Moreover, several works further propose the joint design of multiple components, which is presented in Section~\ref{sec:comprehensive}. Finally, some emerging topics, such as GNNs-based recommendation, multi-modality recommendation are presented in Section~\ref{emerging}.
The frequently used notations are shown in Table~\ref{Table:notations}.

\begin{table}[htbp]

\footnotesize

\caption{Frequently used notations.}
 \resizebox{0.7\textwidth}{!}{
\begin{tabular}{@{}cc@{}}
\toprule
\textbf{Notations}         & \textbf{Descriptions}     \\ \midrule
$\mathbf{x}_i$ & feature one-hot vector of the $i^{th}$ field  \\
$V_i$      & vocabulary size of the $i^{th}$ field \\
$V$      & vocabulary size of the dataset \\
$\mathbf{E}_i$         & embedding table of the $i^{th}$ field  \\
$\mathbf{E}$         & global embedding table  \\
$f_i$ & the $i^{th}$ feature value\\
$\mathbf{e}_i$ & feature embedding vector of the $i^{th}$ field  \\
$\mathbf{e}$ & concatenated feature embedding vector \\
$d$ & embedding size \\
$m$ & number of feature fields\\
$p$ & number of order \\
$\hat{y}$ & predicted score\\
$y$ & ground truth label\\
$a$ & number of candidate sub-dimensions\\
$b$ & number of feature groups\\
$c$ & number of candidate interaction functions\\
$n$ & number of pre-defined blocks\\

$\boldsymbol{\alpha}$ & parameters of selection gates\\
$\boldsymbol{\phi}$ & parameters of the embedding layer \\
$\boldsymbol{\varphi}$ & parameters of the feature interaction layer \\
$\mathbf{\Theta}$ & architectural parameters of the controller/policy network/selection gates $\boldsymbol{\alpha}$ \\
$\mathbf{\boldsymbol{\Phi}}$ & model parameters, including $\boldsymbol{\phi}$ and $\boldsymbol{\varphi}$ \\

\bottomrule

\end{tabular}
}
\label{Table:notations}
\vspace{-0.2cm}
\end{table}

\section{AutoML for Feature Selection}
\label{sec:feature}

In recommender systems, feature selection aims to select a subset of relevant features for constructing recommendation models. 
In practical online service providers, data is composed of a massive amount of features, including user portraits, item attributes, behavior features, contextual features as well as combinatorial features based on previous feature types. 
However, some of these raw features may be irrelevant or redundant in recommendations, which calls for effective feature selections that can boost recommendation performance, overcome input dimensionality and overfitting, enhance model generalization and interpretability, as well as accelerate model training.

The classic feature selection methods are typically presented in three classes: 1) Filter methods, which select features based only on feature correlations regardless of the model~\cite{hall1999correlation,yu2003feature}; 2) Wrapper methods, which evaluate subsets of features that allow detecting the possible interactions amongst variables~\cite{maldonado2009wrapper}; and 3) Embedded methods, where a learning algorithm performs feature selection and classification simultaneously, such as LASSO~\cite{fonti2017feature} and decision trees~\cite{ke2017lightgbm}.
These methods, however, usually fail in deep learning-based recommender systems with both numerical and categorical features. For instance, filter methods neglect the dependencies between feature selection and downstream deep recommendation models; Wrapper methods must explore $2^m$ candidate feature subspaces, \textit{i.e.}, keep or drop for $m$ feature fields; Embedded methods are sensitive to the strong structural assumptions of deep recommendation models. 

To deal with above issues, AutoML-based methods are utilized to automatically and adaptively select the most compelling feature subset for recommendations.
According to the stage of feature selection, we categorize the research into two groups: \textit{Selection from Raw Features} and \textit{Selection from Generated Features}.

% \begin{figure}[!t]
%     \centering
%     \setlength{\belowcaptionskip}{-0.5cm}
%     \includegraphics[width=0.52\textwidth]{Figure/map.PNG}
%     \caption{\small{AutoML for Recommder System.}}
% % \vspace{-0.3cm}
%     \label{fig:embedding}
% \end{figure}

\subsection{Selection from Raw Features}

According to a survey from Crowdflower, data scientists spend 80\% of their time on data and feature preparation~\cite{schwab2019new}. Therefore, introducing AutoML into raw feature selection can enhance data scientists' productivity significantly and frees them up to focus on real business challenges.

Kroon et al. \cite{kroon2009automatic} innovatively consider the raw feature (factors) selection as a reinforcement learning (RL) problem. To efficiently select the optimal feature subset, authors propose a model-based RL method~\cite{kuvayev1997model}, where agents are learned to describe the environment and are optimized by planning methods such as dynamic programming. In this work, the environment is set as factored Markov decision processes (MDPs)~\cite{li2008knows} so that agents are correspondingly dynamic bayesian networks (DBNs). 
Specifically, this method selects highly relevant features for action generation. The search space is constructed as a two-layer graph, where the nodes in layers are states of factors from two continuous steps, and the edges indicate the dependency relations for the decided action. The model is optimized by \textit{Know what it knows}~\cite{li2008knows}, and the optimal result is achieved with the specific searching approach, such as exhaustive search, greedy search, \textit{etc}. %Action 角度考虑FS?
%\textbf{ the agent learns a model, represented as a dynamic Bayesian network (DBN), of a factored Markov decision process (MDP), deduces a minimal feature set from this network, and efficiently computes a policy on this feature set using dynamic programming methods. Model-based RL \cite{kuvayev1997model} approach is utilized to enhance the sample efficiency, i.e., the samples gathered by the agent are used to approximate a model of the environment from which a policy is computed using planning methods such as dynamic programming. Assuming the problem is a factored MDP, this model can be represented as a DBN that describes rewards and state transitions as a stochastic function of the agent’s current state and action.}
% This paper presents a new approach to feature selection specifically designed for the challenges of reinforcement learning. In our method, 

FSTD \cite{fard2013using} also leverages RL into feature selection with a single agent to hand a large search space of $2^m$ ($m$ is the number of feature fields), where each candidate is a possible feature subset. %Two variants of FSTD, based on Filter and Wrapper, are devised to select the final feature subset.
To be specific, FSTD regards the selected feature subsets as states of the environment and constructs the environment as graphs to alleviate the search complexity. The action is to add a specific feature, and the reward is computed based on the difference between two consecutive states, named temporal difference (TD) in the literature, which is the performance difference computed based on GaussianSVM \cite{GaussianSVM}. 
The average TD is calculated for each feature as their final score, where the higher scores represent the more predictive features. Based on these scores, FSTD proposes a filter method and a wrapper method for the final feature selection. The filter directly selects several features with the highest scores, and the wrapper recalls the scores achieved by RL episodes and selects the result of the best episode as the selected result. %, which is an approximate result. 
For evaluation efficiency, FSTD conducts the sample selection before selecting features since the dataset with fewer samples requires less computation during the repeated model training for the reward generation.
% State 角度?
%any subset of features is considered as a state and an action is such as taking a new feature in this state, and passing it to a new state. To indicate the influence of taking a feature as an action, the difference between rewards of two consecutive states is employed, and the average of collected rewards for that feature in several iterations is considered as its final score. In the case that a dataset has too many samples, FSTD runs an instance reduction approach at the beginning and pick up only the selected instances for evaluation.

To limit the searching complexity, MARLFS~\cite{liu2019automating,liu2021automated} reformulates the feature selection as a multi-agent reinforcement learning problem. It is distinguished from previous methods that MARLFS assigns an agent to each feature, actions of which are to select or deselect their corresponding features. The state of the environment is constructed based on the value matrix of the selected feature fields, which contains specific feature values of all samples. MARLFS first calculates the overall reward based on the final performance achieved with the selected features and feature mutual information. Then, it attributes the reward to agents that decide to select the corresponding features, while irrelevant agents would receive zero rewards. The framework is optimized by deep Q-network \cite{kaelbling1996reinforcement}, and sample selection based on the gaussian mixture model \cite{reynolds2009gaussian} is also conducted for efficiency. Further efforts attempt to reduce the computations and enhance the selection quality by learning with external knowledge~\cite{fan2020autofs} or reducing the number of agents~\cite{zhao2020simplifying,fan2021autogfs}. 

% Each feature is assigned an agent, and then all feature agents maintain to select or deselect the corresponding feature simultaneously. 
%It assigns an agent to each feature, the actions of these feature agents are to select or deselect their corresponding features, and the state of environment is characteristics of the selected feature subspace. 
%To select sufficient high-quality samples and avoid low-quality samples, MARLFS develops a gaussian mixture model (GMM) based generative rectified sampling strategy. Specifically, MARLFS first trains a GMM with high-quality samples. The trained GMM is then used to generate a sufficient number of independent samples from different mixture distribution components for reinforcement learning. 

Due to the intrinsically low sample efficiency, RL-based methods are still difficult to be integrated into real-world recommender systems with large-scale user-item interactions. To this end, AutoField~\cite{wang2022autofield} is 
proposed for practical recommendations, where the search space is relaxed to be continuous by allocating two variables to control the selection of each feature. To be specific, for an input dataset with $m$ feature fields, there are two choices for each feature field, \textit{i.e.}, ``selected'' or ``dropped''. Motivated by~\cite{liu2018darts}, AutoField defines the search space as a directed acyclic graph with $m$ parallel nodes that stand for $m$ feature fields respectively. Each node is a 2-dimensional vector containing two parameters, \textit{e.g.}, $(\alpha_i^1, \alpha_i^0)$ for $i^{th}$ node represents for ``selected'' and ``dropped''. 
% In other words, there are $2\times N$ parameters in total for the $N$ feature fields, which control the behaviors of the feature selection process. 
In other words, $\alpha_i^1$ is the probability of selecting a feature field and $\alpha_i^0$ is that of dropping the feature field; thus having $\alpha_i^1 + \alpha_i^0 = 1$. Recalling the search result of DARTS (Equation (\ref{eq:darts_softmax})), the feature selection result could be formulated as:
\begin{equation}
\begin{aligned}
\mathbf{\hat e}_i &= \alpha_i^1\times 1 \times \mathbf{e}_i + \alpha_i^0\times 0 \times \mathbf{e}_i, \\
\mathbf{\hat e} &= [\mathbf{\hat e}_1,\dots,\mathbf{\hat e}_m],
\end{aligned}
\end{equation}
where $\mathbf{e}_i$ and $\mathbf{\hat e}_i$ is the original feature embedding and selection result for the $i^{th}$ feature field respectively, and $\mathbf{\hat e}$ is the concatenated feature embeddings. The number ``1'' and ``0'' are used to keep the original embedding or convert it to a all-zero vector. The architecture parameters for the feature selection, \textit{i.e.}, $\{(\alpha_i^1, \alpha_i^0)\}_{i=1\dots,m}$, are optimized by the gradient descent. %, and the evaluation strategy is also similar to DARTS~\cite{liu2018darts}.
AutoField finally obtains selected features by dropping feature fields with larger $\alpha_i^0$.

Considering that the importance of different feature fields is not always the same for all user-item interactions (data samples), another gradient-based automated feature selection framework, AdaFS~\cite{lin2022adafs}, devices an adaptive feature selection to enhance the recommendation quality. The search space is similar to AutoField. However, AdaFS provides both hard and soft feature selection approaches, where the soft feature selection means that weights are attributed to feature embeddings without dropping any feature field for prediction, while the hard feature selection abandons redundant feature fields. To achieve the adaptive feature selection, AdaFS produces feature weights $\alpha_i^j$ by a trainable controller:
\begin{align}
    [\alpha^j_1,\dots,\alpha^j_m] = \mathrm{Controller}(\mathbf{e}^j),
\end{align}
where $\mathbf{e}^j$ is the concatenated feature embedding for the $j^{th}$ sample. $[\alpha^j_1,\dots,\alpha^j_m]$ are specific feature importance of all feature fields for the $j^{th}$ sample. It is noteworthy that an additional superscript $j$ indicating a specific data sample is necessary since AdaFS conducts adaptive feature selection, \textit{i.e.}, attributing different feature importance according to input data samples. Next, AdaFS produces the selection result as:
\begin{align}
     \mathbf{\hat e}^j &= [\alpha_1^j\mathbf{e}_1^j,\dots,\alpha_m^j\mathbf{e}_m^j],
\end{align}
where $\mathbf{\hat e}^j$ is the concatenated selection result for the $j^{th}$ sample. There are zeros in $\alpha^j_i$ if AdaFS conducts the hard selection. The parameter optimization and architecture evaluation follows DARTS~\cite{liu2018darts}. 
% Afterward, the selected features can be evaluated on the validation dataset by gradient descent.

\noindent\textbf{Insight}.
1) Reinforcement learning methods consider the problem of feature selection as a Markov decision process, in which an agent seeks a control policy for an unknown environment given only a scalar reward signal as feedback. 
The training efficiency can be improved via model-based RL \cite{kroon2009automatic}, multi-agent RL~\cite{liu2019automating,liu2021automated}, external knowledge~\cite{fan2020autofs} and reducing the number of agents~\cite{zhao2020simplifying,fan2021autogfs};
2) Gradient-based approaches are more practical to real-world recommender systems with large-scale user-item interactions with their efficiency and simplicity~\cite{wang2022autofield,lin2022adafs}.

\subsection{Selection from Generated Features}
%\zxy{should include AFGSL~\cite{wu2021afgsl} in this subsection? }
In addition to selecting informative features from the raw feature set, some works learn to discover and generate beneficial combinatorial features (\textit{i.e.}, cross features), including categorical and statistical features. 
Cross features are the results of feature crossing, which integrates two or more features for a more predictive feature. For instance, the cross feature `$\text{Age} \otimes \text{Gender}$' might be more effective for the movie recommendation than single features `Age' and `Gender' since the cross feature could provide more fine-grained information. In addition, cross features could also contribute to introducing non-linearity to linear data, and explicitly generated cross features are more interpretable~\cite{autocross}. However, enumerating all cross features for the model construction would downgrade the recommendation quality and require too much computation because of redundant and meaningless cross features, especially when high-level cross features are considered. Consequently, automated feature crossing is highly desirable~\cite{glider,autocross,aefe}.

%Feature crossing represents the co-occurrence of features, which may be highly correlated with the target label. For example, the cross feature `\code{job $\otimes$ company}' indicates that an individual takes a specific job in a specific company, and is a strong feature to predict one's income.
%Feature crossing also adds nonlinearity to data, which may improve the performance of learning methods. For instance, the expressive power of linear models is restricted by their linearity, but can be extended by cross features. %~\cite{rosales2012post,chapelle2015simple}.
%Last but not least, explicitly generated cross features are highly interpretable, which is an appealing characteristic in many real-world businesses, such as medical treatment and fraud detection.

GLIDER~\cite{glider} utilizes the gradient-based neural interaction detector to detect generic non-additive and high-order combinatorial features efficiently, whose search space is $2^{2^m}$. The detected features are evaluated by a linear regression model and trained from scratch. Specifically, GLIDER first disturbs the feature values by LIME~\cite{ribeiro2016should}, which helps to enhance the model interpretability. Then, the authors compute the importance of cross features based on the partial gradient of the Neural Interaction Detection (NID)~\cite{tsang2017detecting} outputs, where the cross feature search space is set as graphs. GLIDER further encodes selected cross features via Cartesian product and conducts feature truncation for efficiency. This method could exclude repeated training for the model evaluation since their gradient-based NID returns the feature importance with a converged MLP.

%To be specific, GLIDER first detects  feature interactions that span globally across multiple data-instances from a source recommender model, then explicitly encodes the interactions in a target recommender model, both of which can be black-boxes. GLIDER achieves this by first utilizing our ongoing work on Neural Interaction Detection (NID)~\citep{tsang2017detecting} with a data-instance perturbation method called  LIME~\citep{ribeiro2016should} over a batch of data samples. GLIDER then explicitly encodes the collected global interactions into a target model via sparse feature crossing.

Similarly, AutoCross~\cite{autocross} searches useful high-order cross features by transferring the original space to a tree-structured space, reducing the search space from $2^{2^m}$ to $(\mathcal{C}_m^2)^k$, where $k$ is the expected number of cross features. Then, a greedy-based beam search~\cite{beam_search} is performed to prune unpromising branches to improve the efficiency. The feature set evaluation is achieved by field-wise logistic regression approximately and trained from scratch. In detail, the feature set is progressively expanded, where new features are produced by feature crossing on the original feature set and are selected by beam search. This procedure would terminate if the best performance is achieved or the maximal run-time and feature number are reached.

%From the users' perspective, AutoCross is a black box that takes as input the training data and feature types (i.e., categorical, numerical, time series, etc.), and outputs a feature producer. Inside the black box, the data will first be preprocessed, where hyper-parameters are determined, missing values filled and numerical features discretized. Afterwards, useful feature sets are iteratively constructed in a loop consisting of two steps: 1) \textit{feature set generation}, where candidate feature sets with new cross features are generated; and 2) \textit{feature set evaluation}, where candidate feature sets are evaluated and the best is selected as a new solution. This iterative procedure is terminated once some conditions are met.

To discover useful statistical features from the raw feature set, AEFE~\cite{aefe} designs second-order combinatorial features search space with size $2^{\mathcal{C}_m^2Q+Q}$, where $Q$ is the number of pre-defined constructed cross features. The search space contains both the automated feature crossing process $2^{\mathcal{C}_m^2Q}$ and generated feature selection $2^Q$. AEFE applies iterative greedy search for the both feature construction and selection, where the search result is evaluated by the model performance improvement. For the generated features, AEFE adopts all three kinds of methods (filter, embedded, wrapper) in a row. The filter first drops generated features with low variance, which implies that corresponding features are not distinguishable. Then, an embedded method with recommendation models that are capable of generating feature importance (\textit{e.g.}, GBDT) selects essential features according to the produced weights. A wrapper finally selects the optimal features according to the model feedback, \textit{i.e.}, performance improvement. For the performance evaluation, models are directly trained from scratch, and the data sampling strategy is applied before the AEFE is operated for efficiency.

\noindent\textbf{Insight}.
1) The combinatorial features can bring great precision improvement to prediction. They are highly interpretable, which is helpful for digging deep into the underlying relationship of the data; 
2) Due to the large search space and heavy storage pressure, AutoML tools and techniques are utilized to improve the efficiency of selection from generated features.

% Reinforcement learning methods consider the problem of feature selection as Markov decision process, in which an agent seeks a control policy for an unknown environment given only a scalar reward signal as feedback. 
% The training efficiency can be improved via model-based RL \cite{kroon2009automatic}, multi-agent RL~\cite{liu2019automating,liu2021automated}, external knowledge~\cite{fan2020autofs} and reducing the number of agents~\cite{zhao2020simplifying,fan2021autogfs}.
% 2) Gradient-based approaches are more the practical to real-world recommender systems with large-scale user-item interactions with its efficiency and simplicity.
\section{AutoML For Feature Embedding}
\label{sec:embedding}
\begin{figure}[!t]
    \centering
    \setlength{\belowcaptionskip}{-0.5cm}
    \hspace*{-6.6mm}
    \includegraphics[width=0.75\textwidth]{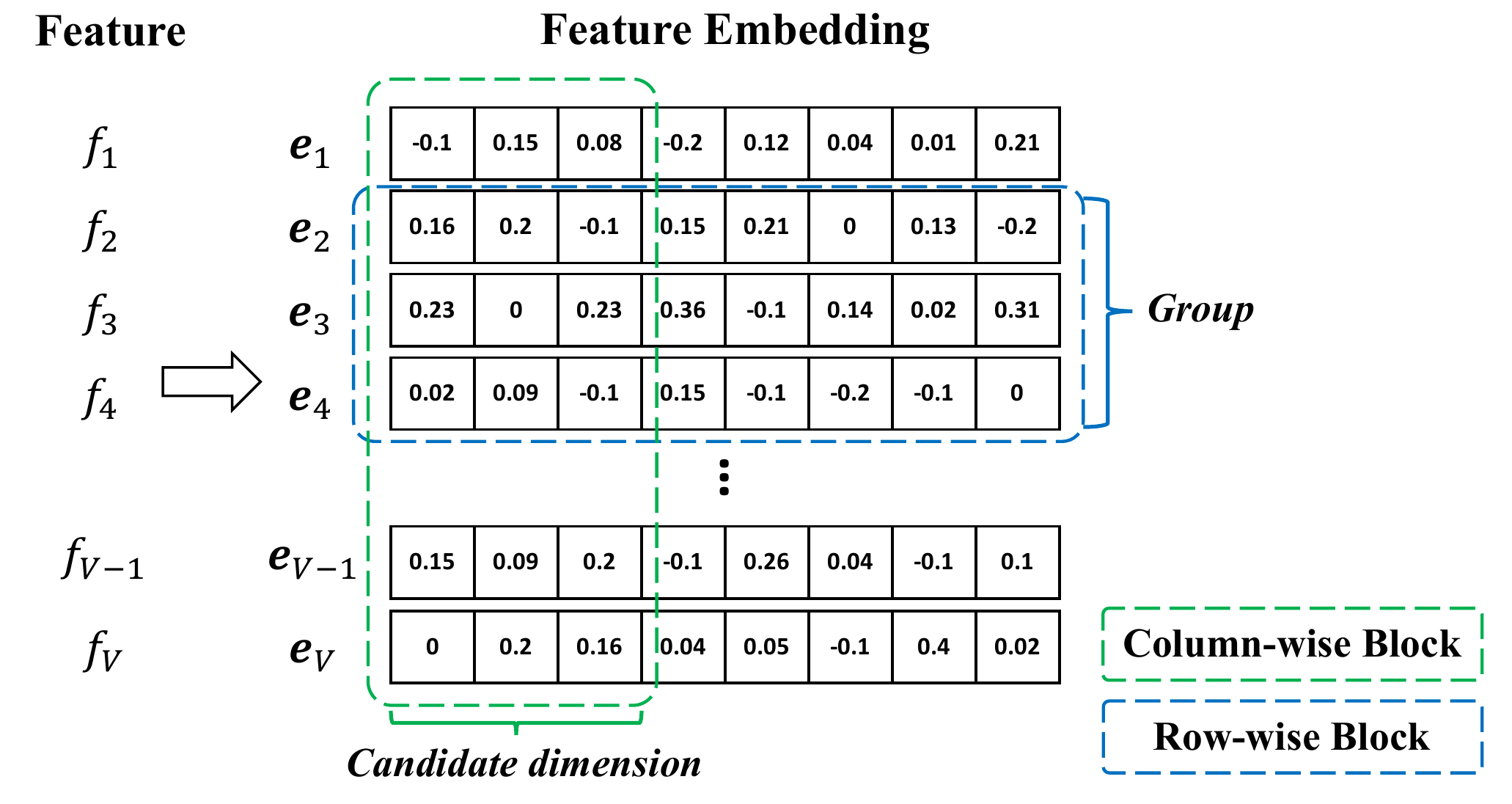}
    \caption{AutoML for feature embedding.}
% \vspace{-0.3cm}
    \label{fig:embedding}
\end{figure}

Different from  Computer Vision (CV) and Natural Language Processing (NLP), the input features used in recommender systems are extremely sparse and high-dimensional. %, thus preventing direct input to the neural networks. 
To tackle this problem, neural network-based models leverage a feature embedding layer to map the high-dimensional features into a low-dimensional latent space. %\textcolor{red}{The most widely-used method for categorical features is embedding look-up operation, which embeds each feature $f_i$ with a dense vector $\mathbf{e}_{i}$ via an embedding table $\mathbf{E}\in \mathbb N^{V\times d}$, whose rows and columns are the vocabulary size $V$ and pre-defined embedding size $d$ respectively. The embedding look-up process is shown in Figure~\ref{fig:embedding}. }
Specifically, for a feature field, each feature value $f_i$ is assigned with a dense embedding vector $\mathbf{e}_{i}$ and all embedding vectors are stored in an embedding table $\mathbf{E}\in \mathbb R^{V\times d}$, where $V$ and $d$ are are the vocabulary size and pre-defined embedding size, respectively. As shown in Figure~\ref{fig:embedding}, based on the embedding table $\mathbf{E}$, we can obtain the embedding vectors through the embedding look-up process.

Feature embedding is the cornerstone of the DRS as the number of parameters in DRS is heavily concentrated in the embeddings and the subsequent feature interaction component is constructed on the basis of feature embeddings. The feature embedding layer not only directly affects storage capacity and online inference efficiency~\cite{autodis}, but also has a non-negligible effect on the prediction accuracy.
To improve the prediction accuracy and save storage space, some AutoML-based solutions are proposed to dynamically search the embedding sizes for different features. The intuition behind is that assigning high-dimensional embeddings for high-frequency features can improve model capacity while low-dimensional embeddings for low-frequency features contribute to preventing overfitting due to over-parameterizing~\cite{autoemb}. 

According to the different search space, these solutions can be divided as into these categories: \textit{Full Embedding Search}, \textit{Column-based Embedding Search}, \textit{Row-based Embedding Search}, and \textit{Column\&Row-based Embedding Search}, as in Figure~\ref{fig:embedding}. Besides, \textit{Combination-based Embedding Search} methods are also presented.
The comparison of these work is presented in Table~\ref{summary_embedding}.

% The feature embedding is critical to DLRMs for two reasons: (1) Feature embedding is the cornerstone of the subsequent feature interaction modeling and has high impact on the prediction performance; (2) The number of parameters in deep recommender models is heavily concentrated in the embedding component, which directly affects the storage capacity and online inference complexity~\cite{autodis}. 
% To improve the prediction accuracy, save storage space and reduce model capacity, some AutoML-based solutions are proposed to dynamically search the embedding sizes for different features. The intuition behind is that assigning high-dimensional embeddings for high-frequency features can improve model capacity while low-dimensional embeddings for low-frequency features contributes to preventing overfitting~\cite{autoemb}. According to whether or not the optimal embedding size is searched for each feature, these solutions can be divided into two categories: \textit{Single Embedding Search} and \textit{Group Embedding Search}, shown in Figure~\ref{fig:embedding}.

\subsection{Full Embedding Search}
Full embedding search-based methods~\cite{amtl,pep} perform the finest-grained embedding dimension search over the original embedding table $\mathbf{E}$, aiming to search optimal embedding dimension for each feature value $f_i$. The advantages of these methods are: 1) Fully consider the impact of each feature embedding dimension on the prediction result, where high-/low-frequency feature values can be assigned with different embedding dimensions, thus improving prediction accuracy. 2) Some numerous but unimportant low-frequency features can be identified and the storage space of their embeddings can be reduced. However, these methods also contain some drawbacks: 1) The search space is extremely huge given the large vocabulary size $V$ and embedding size $d$, which is hard to efficiently search. 2) The dynamical embedding vectors are different to store in the fix-width embedding table $\mathbf{E}$, thus hard to effectively reduce the storage space.

AMTL~\cite{amtl} develops a soft Adaptively-Masked Twins-based Layer over the embedding layer to automatically select appropriate embedding dimension for each feature value with a $d^V$ embedding search space.
A twins-based architecture is leveraged to avoid the unbalanced parameters update problem due to the different feature frequencies, which acts as a frequency-aware policy network to search the optimum embedding dimensions.
To make the learning process non-differentiable, the discrete searching process is relaxed to a continuous space by temperated softmax~\cite{kd} with Straight-Through Estimator (STE)~\cite{ste} and optimized by gradients.

PEP~\cite{pep} proposes a pruning-based solution by enforcing column-wise sparsity on the embedding table $\mathbf{E}$ with $L_0$ normalization. The search space of PEP is $2^{Vd}$, which is much huger than AMTL and is hard to achieve good solution via gradient-based methods. To avoid setting pruning thresholds manually, inspire by the Soft Threshold Reparameterization~\cite{kusupati2020soft}, PEP utilizes the trainable threshold $s \in \mathbb R$ to prune each element automatically:
\begin{equation}
\label{embedding}
    \hat{\mathbf{E}} = \operatorname{sign}(\mathbf{E})\operatorname{ReLU}(|\mathbf{E}|-\operatorname{sigmoid}(s)),
\end{equation}
where $\hat{\mathbf{E}}$ is the re-parameterized embedding table.
Therefore, the learnable threshold $s$ can be jointly optimized with the model parameters $\boldsymbol{\Phi}$ via gradient-based back-propagation. 

\begin{table}[]
\caption{Summary of feature embedding search.}
\label{summary_embedding}
 \resizebox{\textwidth}{!}{
\begin{tabular}{c|cccccc}
\toprule \toprule
\textbf{Method}  & \textbf{Column-Wise} & \textbf{Row-Wise} & \textbf{Search Space} & \textbf{Multi-Embedding} & \textbf{Search Algorithm} & \textbf{Memory Reduction} \\
\midrule
\textbf{AMTL}~\cite{amtl}    & $\times$          & $\times$             & $d^V$                 & $\times$                 & Gradient                  & $\times$                  \\
\textbf{PEP}~\cite{pep}     & $\times$          & $\times$             & $2^{Vd}$              & $\times$                 & Regularization            & $\times$                  \\
\textbf{AutoEmb}~\cite{autoemb}& $\surd$   & $\times$                      & $d^{V}$              & $\surd$                  & Gradient                  & $\times$                  \\
\textbf{ESAPN}~\cite{esapn}   & $\surd$    & $\times$                    & $a^{V}$               & $\surd$                  & RL                        & $\times$                  \\
\textbf{SSEDS}~\cite{sseds}    & $\times$        & $\surd$              & $2^{md}$               & $\times$                 & Gradient                  & $\surd$                   \\
\textbf{AutoSrh}~\cite{autosrh}       & $\times$      & $\surd$               & $2^{bd}$              & $\times$                 & Gradient                  & $\surd$                   \\
\textbf{AutoDim}~\cite{autodim} & $\surd$           & $\surd$              & $a^{m}$               & $\times$                 & Gradient                  & $\surd$                   \\
\textbf{NIS}~\cite{nis}     & $\surd$           & $\surd$              & $ab \; or \; b^a$             & $\times$                 & RL                        & $\surd$                   \\
\textbf{RULE}~\cite{rule}    & $\surd$           & $\surd$              & $2^{ab}$              & $\times$                 & Evolutionary              & $\surd$                   \\
\textbf{ANT}~\cite{ant}        &       -     &      -       &        $2^{kV}$               &      $\times$    &    Gradient            &            $\surd$       \\
\textbf{AutoDis}~\cite{autodis}      &      -      &   -    &     $2^{kV}$           &       $\times$       &        Gradient    &  $\surd$\\
                 \bottomrule \bottomrule
\end{tabular}}
\end{table}

\noindent\textbf{Insight}.
1) Full embedding search methods aim to searching optimal embedding dimension for each feature value, facing huge search space (\textit{e.g.}, AMTL ($d^V$) and PEP ($2^{Vd}$)) and impeding the search efficiency; 2) To facilitate the search procedure, several approaches are proposed to shrink the search space, which can be categorized into three kinds: column-based, row-based, and column$\&$row-based.

\subsection{Column-based Embedding Search}
\label{column-based}
Full embedding search methods search the optimal dimensions over the whole embedding table, which is unnecessary because slight differences in embedding dimensions are difficult to capture. 
Therefore, to reduce the search space, a common solution is to divide the original embedding dimension $d$ into several \textit{\textbf{column-wise}} sub-dimensions (\textit{e.g.}, slicing the original dimension $d = 64$ into $\mathcal{D} = \{2,4,8,16,32,64\}$ 6 candidate sub-dimensions). Therefore, each feature can be assigned an appropriate sub-dimension from the candidate dimension set $\mathcal{D}$.

AutoEmb~\cite{autoemb} and ESAPN~\cite{esapn} focus on searching suitable embedding dimensions for different users and items, and reduce the search space to $d^V$ and $a^V$ respectively, where $a$ is the number of candidate sub-dimensions, greatly shrinking the search space compared with PEP and AMTL.
For both AutoEmb and ESAPN, a set of various embedding sub-dimensions $\mathcal{D}$ is pre-specified. Then, each feature value obtains multiple candidate sub-embeddings, that are transformed into the same dimension via a series of linear transform layers. To make the magnitude of the transformed embeddings comparable, batch normalization with Tanh function is performed.
Specifically, AutoEmb leverages two controller networks (architectural parameters $\mathbf{\Theta}$) to decide the embedding dimensions for each user and item separately, and performs a \textit{soft-selection} strategy by summing over the candidate sub-dimensions with learnable weights. 
Instead, ESAPN performs a \textit{hard-selection} strategy and two frequency-aware policy networks (architectural parameters $\mathbf{\Theta}$) serves as automated RL agents to decide whether to enlarge the dimensions under the streaming setting. 
The optimization of both AutoEmb and ESAPN is achieved by a bi-level procedure (Equation~\ref{eq:darts_bilevel}), where the controller/policy architectural parameters $\mathbf{\Theta}$ are optimized upon the validation set, while the model parameters $\mathbf{\boldsymbol{\Phi}}$ are learned on the training set. 
% \begin{equation}
%     \mathop{\min_{\boldsymbol{\Theta}}}\;\mathcal{L}_{val}(\mathbf{\boldsymbol{\Phi}}^*(\boldsymbol{\Theta}), \boldsymbol{\Theta})
%     \quad\quad s.t. \;\mathbf{\boldsymbol{\Phi}}^*(\boldsymbol{\Theta}) = \arg\min_{\mathbf{\boldsymbol{\Phi}}}\mathcal{L}_{train}(\mathbf{\boldsymbol{\Phi}}, \boldsymbol{\Theta}^*)
% \end{equation}

\noindent\textbf{Insight}.
1) Dividing the embedding dimension into column-wise sub-dimensions (\textit{e.g.}, AutoEmb, ESAPN) is conducive to reducing the search space;
2) Gradient-based approaches is a popular search algorithm due to its efficiency and simplicity;
3) Using multiply embedding tables~\cite{autoemb,esapn} to generate several embedding vectors (\textit{e.g.}, AutoEmb, ESAPN) may incur obvious memory overhead, which can be avoid by shared-embeddings~\cite{autodim};
4) Searching dimensions for each feature value will cause variable-length embedding vectors, which are hard to store in the fix-width embedding table and reduce memory actually.

\subsection{Row-based Embedding Search}
\label{row-based}
AutoEmb and ESAPN shrink the search space by dividing the embedding dimension into candidate \textit{\textbf{column-wise}} sub-dimensions.
Another solution is to group the feature values of a field based on some indicators (\textit{e.g.}, frequencies) and assign a \textit{\textbf{row-wise}} group embedding dimension for all the values within the group. In comparison with the full embedding search, row-based embedding search has the following advantages, making it more practical: 1) Aggregating similar feature values into a group and assigning a unified embedding dimension can greatly shrink the search space, making it easier for the search algorithm to explore satisfactory results. 
2) Row-wise embeddings with same dimension can be stored in an equal-width embedding table, thus saving the storage space physically.

A special case is setting the number of groups for a feature field as $b=1$ and searching a global embedding dimension for all the feature values within the field (namely, field-wise embedding dimension search), such as SSEDS~\cite{sseds}. SSEDS proposes a single-shot pruning method and searches optimal dimension for each feature field, which has $2^{md}$ search space. The core idea is to calculate saliency criterion for identifying the importance of each embedding dimension, which is measured by the change of the loss value and is presented as:
\begin{equation}
\label{sseds}
\begin{aligned}
    &\mathop{\min_{\mathbf{E},\boldsymbol{\Phi}}}\;\mathcal{L}(\mathbf{E}\odot \boldsymbol{\alpha},\boldsymbol{\Phi}) \quad\quad s.t. \; \boldsymbol{\alpha}\in\{0,1\}^{\sum_{i=1}^mV_i\times d},\;\; ||\boldsymbol{\alpha}||_0<\kappa||\mathbf{E}||_0\\
    &s_{i,j} = \Delta\mathcal{L}_{i,j}=\mathcal{L}(\mathbf{E}\odot \mathbf{1},\boldsymbol{\Phi}) - \mathcal{L}(\mathbf{E}\odot(\mathbf{1}-\epsilon_{i,j}),\boldsymbol{\Phi}), 
\end{aligned}
\end{equation}
where $\mathbf{E}$ is the pretrained embedding table, $\boldsymbol{\alpha}$ is the architecture parameters and $\kappa\in(0,1]$ is a budget hyper-parameter, $\mathbf{1}\in\mathbb{R}^{\sum_{i=1}^mV_i\times d}$ is an all-1 matrix and $\epsilon_{i,j}\in\{0,1\}^{\sum_{i=1}^mV_i\times d}$ is a binary matrix with 0 everywhere except for the position on $j^{th}$ dimension of the $i^{th}$ feature field. To avoid $md$ forward passes, the saliency criterion $s_{i,j}$ is approximated by the gradients $\Delta\mathcal{L}_{i,j}\thickapprox\partial \mathcal{L}/\partial \boldsymbol{\alpha}_{i,j}$. 
Top-$k$ saliency scores can be retained by adjusting the budget hyper-parameter $\kappa$ and the model will be retrained to save storage and further boost the performance.

To balance the search efficiency and performance, some works split the feature values within a field into multi-groups (\textit{i.e.}, $b>1$) based on the feature frequencies or clustering. AutoSrh~\cite{autosrh} divides the feature values with similar frequencies into $b$ groups, reducing the search space from $2^{Vd}$ into $2^{bd}$. Then, a gate-based soft selection layer $\boldsymbol{\alpha}\in \mathbb{R}^{b \times d}$ with gradient normalization is used to relax the search space to be continuous, which is further optimized via bi-level gradient-based differentiable search (Equation~\ref{eq:darts_bilevel}).
After optimization, the soft selection layer $\boldsymbol{\alpha}$ is applied to the embedding layer and the non-informative embedding dimensions will be pruned with a pre-defined global threshold to obtain a hard solution.
Remarkably, this hard selection strategy via pre-defined fixed threshold may need lots of human effort to tuning.

\noindent\textbf{Insight}.
1) Row-based embedding search methods explore optimal embedding dimension for a group of feature values, shrinking the search space significantly;
2) In comparison with the column-based embedding search methods, row-based embedding search methods conduces to truly saving memory because feature values within a group are assigned with a same embedding dimension and can be stored in a fix-width embedding table.

\subsection{Column\&Row-based Embedding Search}
As mentioned in Section~\ref{column-based} and~\ref{row-based}, column-based and row-based methods make different assumptions to reduce the search space from different perspective.
To further shrink search space and improve search efficiency, several works combine these two methods and reduce the search space significantly.

Similar as SSEDS~\cite{sseds}, AutoDim~\cite{autodim} also belongs to field-wise embedding dimension search, that searches a global embedding dimension for all the feature values within the field.
AutoDim pre-defines several candidate sub-dimensions $\mathcal{D}$ like ESAPN~\cite{esapn} but has a smaller search space, shrinking from $a^V$ to $a^m$.
During the dimensionality search stage, similar as ESAPN, a set of magnitude-comparable candidate embeddings are obtained via transform layer and batch normalization layer with Tanh function. It is noteworthy that, to avoid maintaining multiple sub-embedding tables like AutoEmb and ESAPN, a weight-sharing embedding allocation strategy is proposed to reduce the storage space and increase the training efficiency. Then, AutoDim introduces structural parameters $\boldsymbol{\alpha}$ and the dimension search algorithm is achieved by Gumbel-Softmax~\cite{gumbel_softmax}. The selection probability $p_{i,j}$ of the $i^{th}$ feature field over the $j^{th}$ sub-dimension is:
\begin{equation}
\label{gumbel}
    p_{i,j} = \frac{\mathop{exp}(\frac{\mathop{log}(\alpha_{i,j})+G_j}{\tau})}{\sum_{k=1}^a\mathop{exp}(\frac{\mathop{log}(\alpha_{i,k})+G_k}{\tau})}, 
\end{equation}
where $G$ follows the standard Gumbel distribution $G  = -\mathop{log}(-\mathop{log}(u))$ with $u \sim \mathop{Uniform}(0, 1)$, $\tau$ is a temperature parameter.
The optimization is achieved through the bi-level gradient-based procedure (Equation~\ref{eq:darts_bilevel}). Finally, a parameter re-training stage is utilized to derive the optimal embedding dimension for each feature field and re-train the model parameters $\mathbf{\Phi}$.

NIS~\cite{nis} and RULE~\cite{rule} also reduce the search space significantly from both row-wise and column-wise perspectives, thus containing much smaller search spaces.
Specifically, NIS designs single-size embedding search mode (with $ab$ space) and multi-size embedding search mode (with $b^a$).
The original embedding table $\mathbf{E}$ is spilt into multiple embedding blocks $\mathbf{E}_{i,j}$, where $i\in[1,2,\dots,b], j\in[1,2,\dots,a]$. Then a RL-based controller learns to sample embedding blocks. 
For the single-size embedding pattern, the controller samples one pair $(i, j)$ from the search space; while for the multi-size embedding pattern, the controller makes a sequence of $a$ choices $[(i_1,1),(i_2,2),\dots,(i_a,a)]$. 
Then, inspired by the the A3C~\cite{mnih2016asynchronous}, the controller is trained with the cost-aware reward over the validation dataset, which takes both optimization objective and training cost into consideration; while the recommendation model is updated over the training dataset.

Similarly, for the on-device recommendation scenario, RULE~\cite{rule} divides the embedding table into multi-blocks and builds a $2^{ab}$ search space. 
To learn expressive embeddings for high-quality recommendation, a embedding pretraining stage is proposed, where Bayesian Personalized Ranking (BPR)~\cite{bpr} with block diversity regularization is performed to optimize all embeddings.
Then the evolutionary search algorithm is proposed to search optimal item embeddings under memory constraint. To improve evaluation efficiency, each sampled sub-structure is evaluated by a performance estimator, which is pretrained over an estimation dataset to a balance the trade-off between prediction confidence and training time.

\noindent\textbf{Insight}.
1) Although it is theoretically optimal to search the optimal dimension for each feature value $f_i$ finely, it poses great challenges to efficient search algorithm. Instead, shrinking the search space in an appropriate manner may result in better performance thanks to adequate search.
2) Reducing the search space from both row-wise and column-wise perspectives attributes to reducing the search space and achieving better results, which becomes the mainstream search method gradually. The evolution of search space from detailed to abstract can lead to higher efficiency.

\subsection{Combination-based Embedding Search}
Besides searching embedding dimension dynamically for each features, learning embeddings via combination adaptively is also a trend for feature representation learning. ANT~\cite{ant} and AutoDis~\cite{autodis} leverage the combination over a set of anchor embeddings (also named meta-embeddings in AutoDis) to represent categorical and numerical feature respectively, building a $2^{kV}$ search space, where $k$ is the number of anchor embeddings. ANT uses a sparse transformation operation to hard-select relevant anchor embedding while AutoDis designs an automatic discretization network to soft-select informative meta-embeddings. 

\noindent\textbf{Insight}.
1) Combination-based embedding learning approach (\textit{e.g.}, ANT, AutoDis) is a novel representational learning method, which 
can efficiently search and greatly save memory usage;
2) Numerical feature representation learning (\textit{e.g.}, AutoDis) is an important but less-explored task.
\section{AutoML for Feature Interaction}\label{sec:fi}
\begin{figure}[!t]
    \centering
    \setlength{\belowcaptionskip}{-0.1cm}
    \hspace*{-6mm}\includegraphics[width=0.95\textwidth]{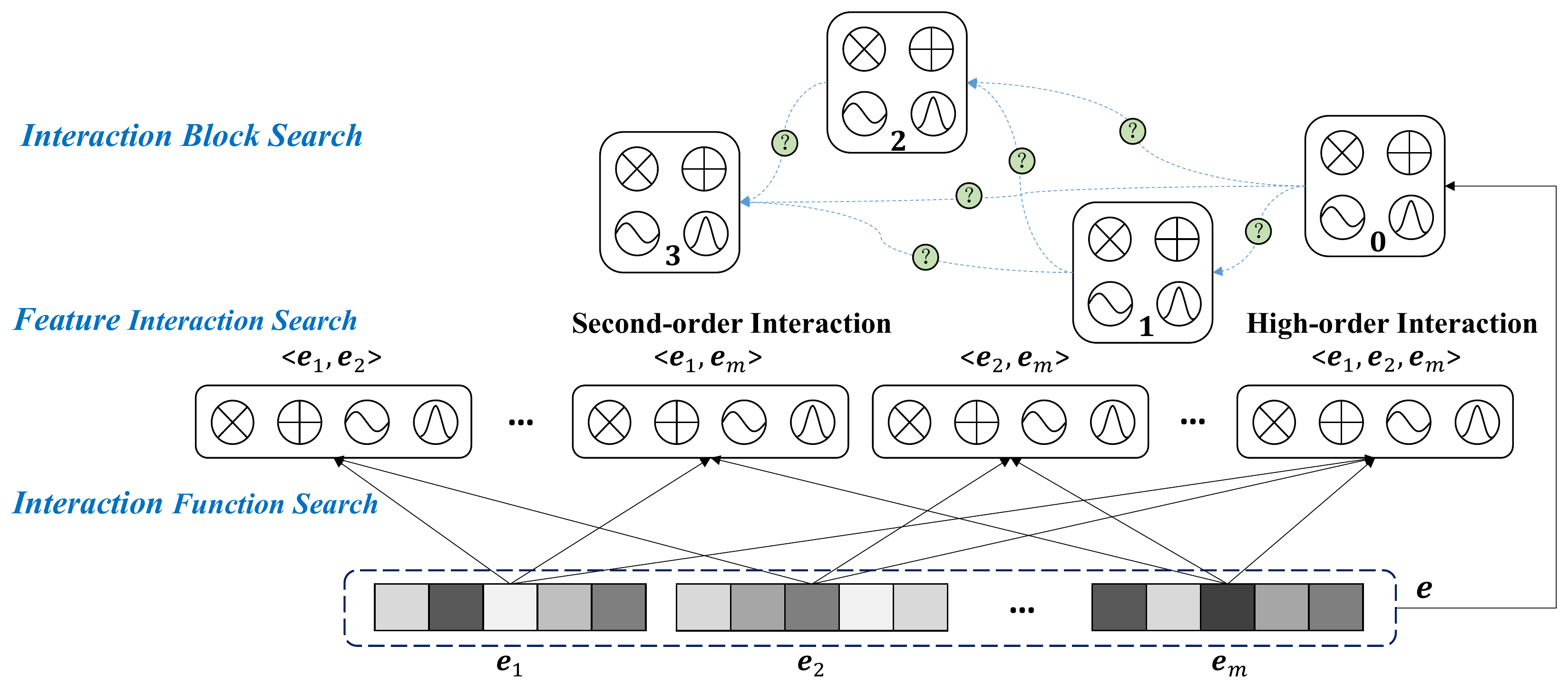}
    \caption{AutoML for feature interaction.}
 \vspace{-0.3cm}
    \label{fig:interaction}
\end{figure}

Effectively modeling feature interactions is one of the most commonly-used approaches for DRS models to improve prediction performance. Recently, plenty of works leverage various operations to capture informative interaction signals explicitly and implicitly, such as inner product (PNN~\cite{pnn}), outer product (CFM~\cite{cfm}), convolution (FGCNN~\cite{fgcnn}) and \textit{etc}. 
As mentioned in Section~\ref{preliminary}, these work can be divided into two kinds: 1) Explicit interaction modeling, mainly factorization-based models~\cite{fm,pnn}. 
However, these works utilize identical interaction functions to model all the feature interactions indiscriminately, which may introduce noisy interactive signals and weaken the effectiveness of modeling. 
2) Implicit high-order interaction modeling. However, designing interaction operations requires a great deal of expert knowledge and it is difficult for these hand-designed interaction operations to achieve consistent performance cross different dataset~\cite{zhu2021open}.
To overcome these issues, some AutoML-based methods are developed to search beneficial feature interactions with optimal interaction function adaptively. These methods can be categorized into three groups depending on the search space: \textit{Feature Interaction Search}, \textit{Interaction Function Search}, and \textit{Interaction Block Search}, as shown in Figure~\ref{fig:interaction}. The comparison of these work is presented in Table~\ref{summary_interaction}.

\subsection{Feature Interaction Search}
Factorization models (\textit{e.g.}, FM~\cite{fm} and PNN~\cite{pnn}) focus on capturing interactive signals among feature subsets, resulting in lots of feature interactions. For the $p$-order feature interaction modeling, $\mathcal{C}_m^p$ feature interactions will be involved, which may contain noisy signals. 
To search beneficial feature interactions for enriching information, some AutoML-based works design high-order feature interactions search space and leverage search algorithms (mainly gradient-based search algorithm) to derive feature interactions automatically.

AutoFIS~\cite{autofis} identifies and selects important feature interactions by enumerating all the feature interactions and introducing a set of architecture parameters ``gates'' $\boldsymbol{\alpha}$ to indicate the importance of individual feature interactions, facing $2^{\mathcal{C}_m^2}$ search space even in second-order feature interactions, namely:
\begin{equation}
\label{autofis}
    \sum_{i=1}^m{\sum_{j>i}^m{\alpha_{ij}f_{\boldsymbol{\varphi}}(\mathbf{e}_i, \mathbf{e}_j)}} = \sum_{i=1}^m{\sum_{j>i}^m{\alpha_{ij}\left \langle \mathbf{e}_i, \mathbf{e}_j \right \rangle}}.
\end{equation}
To decouple the estimation of feature interaction importance $\alpha_{ij}$ with the feature interaction signal $f_{\boldsymbol{\varphi}}(\mathbf{e}_i, \mathbf{e}_j) = \left \langle\mathbf{e}_i, \mathbf{e}_j\right \rangle$, a batch normalization layer is deployed to ease the scale issue. 
During the search stage, the architecture parameters $\boldsymbol{\alpha}$ are optimized by gradient descent with GRDA optimizer~\cite{grda} to get a sparse solution automatically. Besides, AutoFIS utilizes a one-level optimization procedure by optimizing architecture parameters $\boldsymbol{\alpha}$ jointly with model parameters $\mathbf{\Phi}$. 
After the search stage, the optimal feature interactions with nonzero $\boldsymbol{\alpha}$ are obtained, and the model will be retrained to fine-tune the parameters. 
However, the disadvantages of AutoFIS is obvious. When searching high-order feature interactions, the search space is extremely huge, resulting in low search efficiency.

To solve the \textit{\textbf{efficiency-accuracy}} dilemma, AutoGroup~\cite{autogroup} proposes automatic feature grouping, reducing the $p^{th}$ order search space from $2^{\mathcal{C}_m^p}$ to $2^{gm}$, where $g$ is the number of pre-defined groups. During the automatic feature grouping stage, the discrete search space is relaxed to continuous by introducing structural parameters $\alpha_{i,j}^p$, which indicates the selection probability of feature $f_i$ for the group $s_j^p$. Then to make the whole optimization procedure differentiable, the derivative is approximated by introducing the Gumbel-Softmax trick~\cite{gumbel_softmax}, which is presented in  Equation (\ref{gumbel}). AutoHash~\cite{autohash} shares a similar idea with AutoGroup to reduce high-order search space by the hashing function.

\begin{table}[]
\caption{Summary of feature interaction search.}
\label{summary_interaction}
 \resizebox{0.8\textwidth}{!}{
\begin{tabular}{c|cccccc}
\toprule \toprule
\textbf{Method}  & \textbf{Search Category}  & \textbf{Search Space} & \textbf{Order} & \textbf{Search Algorithm} \\
\midrule
\textbf{AutoFIS}~\cite{autofis}    & Feature Interaction     & $2^{\mathcal{C}_m^p}$             & second-order                  & Gradient                   \\
\textbf{AutoGroup}~\cite{autogroup}     & Feature Interaction         & $2^{gm}$             & high-order        & Gradient                 \\
\textbf{PROFIT}~\cite{profit} & Feature Interaction       & $mR$              & high-order        & Gradient                         \\
\textbf{FIVES}~\cite{fives}   & Feature Interaction        & $2^{m^2}$              & high-order     & Gradient                                 \\
\textbf{$L_0$-SIGN}~\cite{l0sign} & Feature Interaction        & $2^{m^2}$              & second-order     & Gradient                                 \\
\textbf{BP-FIS}~\cite{chen2019bayesian}        &     Feature Interaction    &    $u2^{\mathcal{C}_m^p}$      &        second-order        &   Bayesian          &             \\
\textbf{SIF}~\cite{sif}      &   Interaction Function   &   $\mathcal{C}_c^n$    & second-order   &        Gradient   \\
\textbf{AutoFeature}~\cite{autofeature} & Interaction Function          & $c^nn{\mathcal{C}_m^p}$              & second-order           & Evolutionary                        \\
\textbf{AOANet}~\cite{aoanet}    & Interaction Function          & $2^{|(i,j)\; pairs|}$             & high-order            & Gradient                   \\
\textbf{AutoCTR}~\cite{autoctr}     & Interaction Block         & $c^{\mathcal{C}_n^2}$              & high-order                      & Evolutionary                       \\
\textbf{AutoPI}~\cite{autopi}    & Interaction Block        & $c^{\mathcal{C}_n^2}$              & high-order                  & Gradient                 \\
                 \bottomrule \bottomrule
\end{tabular}}
\end{table}

Although AutoGroup and AutoHash improve the high-order interaction search efficiency via feature grouping and hashing, they ignore the \textit{\textbf{order-priority}} property ~\cite{profit}, which reveals that the higher-order feature interactions quality can be relevant to their de-generated low-order ones, and lower-order feature interactions are likely to be more vital compared with higher-order ones. To reduce the architecture parameters and search costs in high-order feature interaction searching, PROFIT~\cite{profit} distills the $p^{th}$ order search space from $2^{\mathcal{C}_m^p}$ to $mR$ by decomposing the structural parameters $\boldsymbol{\alpha}$ into order-wise low-rank tensors $\{\boldsymbol{\beta}^1,\boldsymbol{\beta}^2,\dots,\boldsymbol{\beta}^p\}$ approximately:
\begin{equation}
\label{profit}
    \boldsymbol{\alpha} \thickapprox \sum_{r=1}^R \underbrace{\boldsymbol{\beta}^r \circ \dots \circ \boldsymbol{\beta}^r}_{\boldsymbol{\beta}^r \; \text{repeats} \; p \; \text{times}} ,
\end{equation}
where $\boldsymbol{\beta}^r \in \mathbb{R}^{1 \times m}$ and $R$ is a positive scalar.
Then to ensure the order-priority property, a progressive search algorithm based on the gradient is proposed to search high-order feature interactions order-by-order. Specifically, when searching $p^{th}$ order interactions, the architecture parameters with order lower than $p$ are fixed, \textit{i.e.}, $\{\boldsymbol{\beta}^1,\boldsymbol{\beta}^2,\dots,\boldsymbol{\beta}^{p-1}\}$. Finally, the top-{$k$} important interactions in each order is reserved and the model will be retrained to ensure prediction accuracy and model efficiency.

Similar to PROFIT~\cite{profit}, FIVES~\cite{fives} regards the original features as a feature graph conceptually and models the high-order feature interactions by a Graph Neural Network (GNN) with layer-wise adjacency matrix, so that the $p^{th}$ order search space is reduced from $2^{\mathcal{C}_m^p}$ to $2^{m^2}$. Then, FIVES parameterizes the adjacency matrix $\mathbf{A} \in \mathbb{R}^{p\times m \times m}$ and makes each layer $\mathbf{A}^{(p)}$ depend on the previous layer $\mathbf{A}^{(p-1)}$, so that the order-priority property can be kept. To make the search process more efficient, a soft adjacency matrix $\mathbf{A}^{(p)}$ can be obtained by:
\begin{equation}
\label{fives}
\begin{aligned}
    \mathbf{A}^{(p)} & \triangleq (\mathbf{D}^{(p-1)})^{-1} \psi(\mathbf{A}^{(p-1)}) \sigma(\mathbf{H}^{(p)}), \\
    \mathbf{A}^{(0)} & \triangleq \mathbf{I} \quad \mathop{and} \quad \mathbf{H}^{(0)} \triangleq \mathbf{I},
\end{aligned}
\end{equation}
where $\mathbf{D}^{(p-1)}$ is the degree matrix of $\mathbf{A}^{(p-1)}$, $\psi(\cdot)$ is a binarization function with a tunable threshold and $\mathbf{H} \in \mathbb{R}^{p\times m \times m}$ is the learnable parameters. Finally, the hard binary decision can be re-scaled by $\mathbf{A} \leftarrow \sigma(\frac{\mathop{log}[\mathbf{A}/(1-\mathbf{A})]}{\tau})$.
$L_0$-SIGN~\cite{l0sign} shares the similar idea of modeling feature interactions via GNN. The difference is that $L_0$-SIGN only takes the second-order interactions into consideration and generates a $2^{m^2}$ search space. Then, a MF-based edge prediction function instantiated as a neural network with $L_0$ activation function is utilized to search beneficial feature interactions. Therefore, the sparse solution can be achieved by the $L_0$ regularization. Based on the detected interactions, graph information aggregation and graph pooling operations are performed to obtain final prediction.

The above-mentioned works search beneficial feature interactions for all users non-personally, which overlooks the individuality and personality of the user's behavior. To provide personalized selection of second-order feature interaction, BP-FIS~\cite{chen2019bayesian} designs a personalized search space with size $u2^{\mathcal{C}_m^2}$, where $u$ is the number of users. Specifically, BP-FIS proposes bayesian personalized feature interaction selection mechanism under the Bayesian Variable Selection (BVS)~\cite{tibshirani1996regression} theory by forming a Bayesian generative model and deriving the Evidence Lower Bound (ELBO), which can be optimized by an efficient Stochastic Gradient Variational Bayes (SGVB) method.

\noindent\textbf{Insight}.
1) Feature interaction search based methods focus on searching beneficial low-/high-order interactive signals for factorization models;
2) For high-order interaction search, different approaches are proposed to reduce the search space, such as feature grouping~\cite{autogroup}, hashing~\cite{autohash}, tensor decomposition~\cite{profit}, and graph aggregation~\cite{fives};
3) Gradient-based search algorithm is dominant in this task due to the high efficiency;
4) Personalized feature interaction search will incur a much huger search space, bringing challenges to efficient search algorithms.

\subsection{Interaction Function Search}
Although feature interaction search methods are dedicated to exploring beneficial feature interactions, they model these interactive signals with a globally unified interaction function (\textit{e.g.}, inner product or weighted sum).
However, plenty of human-designed interaction functions (\textit{e.g.,} inner/outer product, plus, minus, max, min, concat, conv, MLP, \textit{etc.}) have been applied to various recommendation models.
As suggested by PIN~\cite{pin}, different feature interactions are suitable for different interaction functions. Therefore, searching optimal interaction functions $f_{\boldsymbol{\varphi}}(\cdot)$ for different feature interactions contributes to better extracting informative interactive signals.

SIF~\cite{sif} automatically devises suitable interaction functions for collaborative filtering (CF) task with two fields (user id $\mathbf{e}_u$ and item id $\mathbf{e}_v$), which consists of micro search space referring to element-wise MLP and macro search space including $c = 5$ pre-defined operations $\mathcal{O}$ (\textit{i.e.}, \texttt{multiply}, \texttt{plus}, \texttt{min}, \texttt{max}, and \texttt{concat}), resulting $\mathcal{C}_c^n$ search space, where $n$ is the number of pre-defined search blocks. 
A bi-level gradient-based search algorithm is utilized to relax the choices among operations in a continuous space. 
\begin{equation}
\label{sif}
    f_{\boldsymbol{\varphi}}(\mathbf{e}_u, \mathbf{e}_v) = \sum_{i=1}^{|\mathcal{O}|} \alpha_i \mathcal{O}_i(\dot{\mathbf{e}}_u, \dot{\mathbf{e}}_v) \; \mathop{with} \; \dot{\mathbf{e}}_u =\mathop{MLP}(\mathbf{e}_u;\mathbf{p}) \; \mathop{and} \; \dot{\mathbf{e}}_v =\mathop{MLP}(\mathbf{e}_v;\mathbf{q}),
\end{equation}
where $\mathbf{p}$ and $\mathbf{q}$ are the parameters of MLP. The architecture parameters $\{\boldsymbol{\alpha}, \mathbf{p}, \mathbf{q}\}$ are optimized over the validation set while the other model parameters are optimized over the training set iteratively.

AutoFeature~\cite{autofeature} extends the interaction functions search to multi-field high-order scenarios by utilizing micro-networks with different architectures to model feature interactions. Supposed $n$ is the number of pre-defined search blocks for interaction functions, the whole search space expands to $c^nn\mathcal{C}_m^p$ for the $p^{th}$ order interactions, including $c = 5$ pre-defined operations: \texttt{add}, \texttt{hadamard-product}, \texttt{concat}, \texttt{generalized-product}, and \texttt{null} (\textit{a.k.a.}, remove this interaction). 
The search process is implemented by an evolutionary algorithm. To make this procedure more efficient, a Naive Bayes Tree (NBTree) is utilized to partition the search space based on the accuracy of sampled architectures, where the leftmost leaf subspace represents the most promising subspace. Then, a sampling strategy based on the Chinese Restaurant Process (CRP) is used to sample subspace, where the top-2 samples as parents are performed crossover and mutation operations to generate an offspring architecture.
Each architecture is trained and evaluated from scratch to obtain the accuracy, which is further used to update the NBTree and CRP.
%and the whole search process continues until the desired accuracy is achieved or the maximum number of steps is reached.

However, the interaction functions of SIF and AutoFeature are artificially specified, which requires high dependence on domain knowledge. To overcome this limitation, AOANet~\cite{aoanet} proposes a generalized interaction paradigm by decomposing commonly-used structures into Projection, Interction and Fusion phase. The interaction and fusion layer can be represented as:
\begin{equation}
\label{aoanet}
\begin{aligned}
\mathbf{Z}_{i,j} & = (\mathbf{e}_i\otimes \mathbf{e}_j) \odot \mathbf{W},\\
\mathbf{P} & = \mathbf{h}^T\sum_{i,j}(\alpha_{i,j}\mathbf{Z}_{i,j}),
\end{aligned}
\end{equation}
where $\mathbf{Z}_{i,j}$ is the interaction of vectors $\mathbf{e}_i$ and $\mathbf{e}_j$ that are the original feature embeddings or latent representations, architecture parameter $\alpha_{i,j} \in \mathbb{R}$ indicates the importance of interaction $\mathbf{Z}_{i,j}$, $\mathbf{W}$ and $\mathbf{b}$ are the trainable parameters. 
% Therefore, the formula of the $i^{th}$ interaction layer $C_i$ is given as: 
% \begin{equation}
% \label{aoanet}
% \begin{aligned}
% C_i & = \{Z_{u,v} | (u,v)\in pairing(B_0,B_{i-1})\},\\
% Z_{u,v} & = (u\otimes v) \odot W_i,
% \end{aligned}
% \end{equation}
The optimization procedure is achieved by a gradient-based method like AutoFIS and some unimportant interactions will be pruned during the retrain stage based on the architecture parameters $\boldsymbol{\alpha}$. Note that the interaction modeling formulation in Equation (\ref{aoanet}) is only second-order with search space $2^{|(i,j)\;pairs|}$. To enable high-order interaction, AOANet stacks multiple interaction and fusion layers for improving capacity.

\noindent\textbf{Insight}.
1) Although searching appropriate interaction functions for different feature interactions helps to improve accuracy, the introduced cost is higher, which hinders the application in high-order scenarios;
2) Generalized interaction function search (\textit{e.g.}, AOANet~\cite{aoanet}) is more efficient than searching over human-designed search space (\textit{e.g.}, AutoFeature~\cite{autofeature}), providing a promising paradigm for high-order interaction function search.

\subsection{Interaction Block Search}
Searching appropriate interaction functions for different feature interactions may bring huge search space and high search overhead, especially for high-order interaction modeling. Therefore, to reduce the search space for high-order feature interaction modeling, another route is to take the original features as a whole and model feature interactions over the whole feature sets implicitly, such as MLP, cross operation in DCN~\cite{wang2017deep}, and compressed interaction network in xDeepFM~\cite{lian2018xdeepfm}. Several AutoML works modularize representative interaction functions in several blocks to formulate a search space, which is categorized as interaction block search.

AutoCTR~\cite{autoctr} designs a two-level hierarchical search space $c^{\mathcal{C}_n^2}$ by abstracting the raw features (including categorical and numerical features) and pre-defined operations ($c=3$, \textit{i.e.}, \texttt{MLP}, \texttt{FM}~\cite{fm}, and \texttt{dot-product}) into $n$ virtual blocks, which are further connected as a Directed Acyclic Graph (DAG). Similar to AutoFeature, AutoCTR utilizes a multi-objective evolutionary algorithm to search the optimal architecture. An architecture evaluation metric considered fitness, age, and model complexity is proposed to generate population. Then a sampling method based on the tournament selection is utilized to select parent. To guide the mutation, a guider is trained to evaluate the model accuracy and select a most promising mutated architecture as offspring.
The architectures are evaluated from scratch and some tricks (\textit{e.g.}, data sub-sampling and warm-start) are used to accelerate the evaluation process.

To further improve computational efficiency, AutoPI~\cite{autopi} utilizes a gradient-based search strategy in a similar search space $c^{\mathcal{C}_n^2}$ and a hierarchical search space with $c=6$ pre-defined operations (\texttt{skip-connection}, \texttt{SENET}~\cite{fibinet}, \texttt{self-attention}, \texttt{FM}~\cite{fm}, \texttt{MLP}, and \texttt{1d convolution}) is designed. AutoPI connects $n$ blocks into a DAG and defines two cells where the interaction cell formulates the higher-order feature interactions and the ensemble cell combines lower-order and higher-order interactions. Then, a bi-level optimization approach is applied to discover optimal architecture after the continuous relaxation.

\noindent\textbf{Insight}.
1) Modeling overall high-order feature interactions over the whole feature sets implicitly can significantly shrink the search space~\cite{autoctr,autopi} in comparison with the explicit high-order interaction function search~\cite{autofeature}, making the search procedure more efficient;
2) Although fine-grained feature interactions is important for recommendation~\cite{guo2017deepfm}, neural networks are still capable of identifying important interactive signals from the original information. Recently, recommendation models that implicitly model high-order feature interactions like DCN~\cite{wang2017deep} and xDeepFM~\cite{lian2018xdeepfm} are shown to be effective. Therefore, interaction block search based methods with more abstract search space may become mainstream gradually due to its efficiency.

\section{AutoML for Model Training}
\label{sec:training}
Besides feature and interaction modeling, the training process is also crucial for designing reliable recommender systems. 
In this section, we present some works that automatically facilitate the model training procedure, including loss function design~\cite{zhao2021autoloss}, parameter transfer~\cite{yang2021autoft}, and model implementation~\cite{autorec}.

In general cases, the loss function is vital for model training. GradNorm~\cite{chen2018gradnorm} and $\lambda$Opt~\cite{lambdaopt} focus on adjusting the coefficients of loss items and optimizing parameters via gradient descent. The difference is that GradNorm aims to balance different losses of multi-task while $\lambda$Opt adjusts the regularization level for different users. 

Later, an adaptive loss function search framework AutoLoss~\cite{zhao2021autoloss} is proposed based on a bi-level gradient-based algorithm with the Gumbel-Softmax trick~\cite{gumbel_softmax}. AutoLoss attributes the most appropriate loss function for each data example by automatically designing loss functions, rather than adjusting coefficients only like the aforementioned works. Specifically, AutoLoss formulates the loss function selection problem with probabilities $\boldsymbol{\alpha}$ as:
\begin{align}
    \mathcal{L}(y,\hat y;\boldsymbol{\alpha}) &= \sum_i \alpha_i \mathcal{L}_i(y,\hat y).\label{autoloss:loss}
\end{align}
% \begin{align}
%     \mathcal{L}(y,\hat y;\boldsymbol{\alpha}) &= \sum_i \alpha_i \mathcal{L}_i(y,\hat y),\label{autoloss:loss}\\
%     \boldsymbol{\alpha}&= \mathrm{Controller}(y,\hat y), \label{autoloss:con}
% \end{align}
%where Equation (\ref{autoloss:loss}) defines the loss function search problem with probabilities $\boldsymbol{\alpha}$.%, Equation (\ref{autoloss:con}) illustrates that $\boldsymbol{\alpha}$ is computed according to different convergence behaviors $(y,\hat y)$.

Rather than adopting searching strategies, AutoLossGen~\cite{li2022autolossgen} designs a loss function generation framework based on reinforcement techniques~\cite{kaelbling1996reinforcement}. It utilizes an RNN controller to generate sequences of basic mathematical operators and corresponding variables. Then, AutoLossGen converts these sequences to loss functions by combining these operators and variables. It is worth noting that AutoLossGen does not require prior knowledge of loss function, \textit{e.g.}, loss function candidates, which is totally different from the aforementioned loss function search frameworks. 

In the scenario of multi-domain recommendation, system designers should figure out which parts of parameters should be frozen when transferring pre-trained models from the source dataset to prevent overfitting on the target dataset. AutoFT designs the search space from two perspectives, field-wise and layer-wise. Field-wise parameters are feature embedding matrices, and layer-wise parameters are feature interaction networks, including cross layers and deep layers in DCN. With the Gumbel-Softmax trick~\cite{gumbel_softmax}, AutoFT~\cite{yang2021autoft} automatically decides whether the embedding of a field and parameters of a layer should be fine-tuned. For a specific block $\mathbf{V}$, AutoFT formulates the fine-tune searching problem as:
\begin{align}
 \mathbf{x}_l = \alpha_l \mathbf{V}_p (\mathbf{x}_{l-1}) +  (1 - \alpha_l)\mathbf{V}_f (\mathbf{x}_{l-1}),
\end{align}
where  $\mathbf{x}_l$ is the output of $l^{th}$ layer and $\alpha_l$ is the probability of using pre-trained parameters $\mathbf{V}_p$ or fine-tuned parameters $\mathbf{V}_f$.

Finally, in terms of models' implementation, AutoRec~\cite{autorec}, as the first open-source platform, provides a highly-flexible pipeline for various data formation, tasks, and models in deep recommender systems. Besides, some extensible and modular frameworks are proposed for differentiable neural architecture search, such as DNAS~\cite{krishna2021differentiable} and ModularNAS~\cite{modularnas}. These frameworks significantly alleviate systems designers' burden on designing and implementing novel automated recommendation models.

\noindent\textbf{Insight}.
1) Loss-based optimization methods facilitate recommendation model training via searching optimal loss function automatically, bringing significant results, such as AutoLoss~\cite{zhao2021autoloss} and AutoLossGen~\cite{li2022autolossgen}. However, many aspects of the training process and other parts of DRS are still remained to be explored, \textit{e.g.}, selecting the most suitable optimizer rather than using the a pre-defined one; 
2) We could find that, from GradNorm~\cite{chen2018gradnorm} to AutoLoss~\cite{zhao2021autoloss} and AutoFT~\cite{yang2021autoft}, researchers gradually realize more flexible and efficient methodologies to facilitate model training.  

\section{Comprehensive Search}
\label{sec:comprehensive}
Besides the aforementioned techniques for automating a single component in DRS models (\textit{e.g.}, feature selection, feature embedding, feature interaction or model training), scientists also devise comprehensive components search methods, which is different from works introduced in previous sections. These comprehensive search methods would design hybrid search spaces and search for several key components. We summarize these related works in Table~\ref{tab:com}.

\begin{table}[]
\caption{Summary of comprehensive search}
\label{tab:com}
 \resizebox{0.8\textwidth}{!}{
\begin{tabular}{@{}c|ccccccc@{}}
\toprule
\toprule
\multirow{2}{*}{\textbf{Model}} & \multirow{2}{*}{\textbf{\begin{tabular}[c]{@{}c@{}}Feature \\ Selection\end{tabular}}} & \multirow{2}{*}{\textbf{\begin{tabular}[c]{@{}c@{}}Feature \\ Embedding\end{tabular}}} & \multicolumn{3}{c}{\textbf{Feature Interaction}} & \multirow{2}{*}{\textbf{\begin{tabular}[c]{@{}c@{}}Search \\ Algorithm\end{tabular}}} \\ \cmidrule(lr){4-6}
 &  &  & \textbf{\begin{tabular}[c]{@{}c@{}}Feature \\ Interaction\end{tabular}} & \textbf{\begin{tabular}[c]{@{}c@{}}Interaction \\ Function\end{tabular}} & \textbf{\begin{tabular}[c]{@{}c@{}}Interaction\\ Block\end{tabular}} &  \\ \midrule
\textbf{AIM}~\cite{aim} & $\times$ & $\surd$ & $\surd$ & $\surd$ & $\times$ & Gradient \\
\textbf{AutoIAS}~\cite{autoias} & $\times$ & $\surd$ & $\surd$ & $\surd$ & $\surd$ & RL \\
\textbf{DeepLight}~\cite{deeplight} & $\times$ & $\surd$ & $\surd$ & $\times$ & $\surd$ & Regularization \\
\textbf{UMEC}~\cite{umec} & $\times$ & $\surd$ & $\times$ & $\times$ & \textbf{$\surd$} & Regularization \\
\textbf{OptInter}~\cite{lyu2021memorize} & $\surd$ & $\times$ & $\surd$ & $\times$ & $\times$ & Gradient \\
\textbf{AutoGen}~\cite{autogen} & $\surd$ & $\surd$ & $\times$ & $\times$ & $\surd$ & Gradient \\
\textbf{AMEIR}~\cite{ameir} & $\times$ & $\times$ & $\surd$ & $\times$ & $\surd$ & Random Search \\
\bottomrule
\bottomrule
\end{tabular}
}
\end{table}

In Section~\ref{sec:fi}, we have introduced AutoFIS~\cite{autofis}, which could select effective feature interactions with ``gates'' $\boldsymbol \alpha$. Based on AutoFIS, AIM~\cite{aim} expands the search space to a hybrid one, achieving a comprehensive selection of feature interactions, appropriate interaction functions, and optimal embedding dimensions to integrate a unified framework. Compared with AutoFIS, which can not be extended to high-order interaction search due to high overhead, AIM adopts a progressive search strategy with order-priority property to overcome this issue. To select the ${(p+1)}^{th}$-order feature interaction, AIM interacts raw features ($1^{st}$-order) with the %\footnote{For the simplicity, we use the same notation $m$ for the number of feature fields and the maintained interactions for all orders.}
${p}^{th}$-order feature interactions (only with selected top-$k$). This strategy reduces the search space size for the $p^{th}$-order interaction from $2^{\mathcal{C}_m^p}$ to $2^{km}$. Besides, the ``gates'' $\boldsymbol \alpha$ are extended to search embedding dimensions and interaction functions with search space sizes $2^{md}$ and $2^{kmc}$ respectively ($d$ is the embedding dimension and $c$ is the number of interaction functions). Specifically, for the $(p+1)^{th}$-order interaction, AIM formulates the feature interaction search (FIS) and interaction function search(IFS) problems with ``gates'' $\boldsymbol \alpha$ as:
\begin{align}
FIS_{p+1} &= \sum_{i=1}^m \sum_{j>i}^m \alpha_{ij}f(\mathbf{e}_i,\mathbf{e}_j^p),\label{AIM:FIS} \\ 
IFS_{p+1} &= \sum_{i=1}^m \sum_{j>i}^m \sum_{k=1}^c \alpha_{ij}^kf_k(\mathbf{e}_i,\mathbf{e}_j^p),\label{AIM:IFS}
\end{align}
where $\mathbf{e}_i$ is the $i^{th}$ raw feature, $\mathbf{e}_j^p$ is the $j^{th}$ maintained feature interaction from the selection of $p^{th}$-order interaction, $f(\cdot)$ in Equation (\ref{AIM:FIS}) is a fixed interaction function and $f_k(\cdot)$ in Equation (\ref{AIM:IFS}) is the $k^{th}$ interaction function candidate for IFS. For the embedding dimension search (EDS), AIM organizes the search space with gates $\boldsymbol \beta$:
\begin{align}
    EDS = \sum_{i=1}^d \boldsymbol{1}(\beta_i \neq 0).
\end{align}
To obtain the optimal structure, AIM first searches the optimal feature interactions and corresponding interaction functions. Then, AIM attributes  embedding dimensions to every feature embeddings. AIM optimizes all gates, $\boldsymbol{\alpha}$ for FIS and IFS, and $\boldsymbol{\beta}$ for EDS, via gradients.

AutoIAS~\cite{autoias} also designs an integrated search space for multiple components in DRS, including feature embedding ($a^m$), projection ($a^{\mathcal{C}_m^2}$), second-order feature interaction ($2^{m+\mathcal{C}_m^2}$), interaction function ($c^{\mathcal{C}_m^2}$), as well as the MLP structures. An architecture generator network is trained by policy gradient and used to produce better architectures with dependency, where Knowledge Distillation (KD)~\cite{kd} is performed to enhance consistency among sub-architectures. 

DeepLight~\cite{deeplight} develops an integrated search space for feature embedding ($2^{Vd}$), interaction ($2^{d^2}$), and MLP structures by pruning redundant parameters with $L_2$ penalty.
Resembling DeepLight, UMEC~\cite{umec} develops an integrated search space for both embedding ($2^{Vd}$) and MLP structures. Then the sparsity is achieved by $L_2$ norms, which are further reformulated as a minimax optimization problem and optimized via a gradient-based algorithm.

OptInter~\cite{lyu2021memorize} searches for the combinatorial features (\textit{i.e.}, cross features) together with feature interactions and integrates them into a unified search space. % by regarding feature pairs that have no interaction as a ·`na\"{\i}ve‘’' interaction method. 
All operation candidates are memorized methods (\textit{i.e.}, generation combinatorial features by explicitly viewing interactions as new features and assigning trainable embeddings), factorized methods (\textit{i.e.}, generation feature interaction vectors for original features by implicitly modeling interactions through factorization functions, \textit{e.g.} Haddmard Product), and na\"{\i}ve methods (no interaction). OptInter obtains the final architecture by solving a bi-level optimization problem with the Gumbel-Softmax trick~\cite{gumbel_softmax}.

Existing works are devoted to searching for an optimal model to serve global users. However, searching different sub-models or sub-architectures for different user requests personalized and adaptively is beneficial for industry recommender systems to achieve overall revenue maximization. AutoGen~\cite{autogen} designs a multi-model service for intelligent recommendation and proposes an automated dynamic model generation framework.
AutoGen involves a mixed search space (including input features, feature embeddings and input layers) and leverages an importance-aware progressive training scheme to prevent interference between subnets, thereby generating multiple satisfactory models with different complexities in a short searching time.

The above works only focus on comprehensive components search for non-sequential features. It is noteworthy that AMEIR~\cite{ameir} proposes an automatic behavior modeling, feature interaction exploration and
MLP structure investigation solution, which searches for the optimal structure for both sequential and non-sequential features. For sequential behavior modeling, AMEIR designs a block search space to identify the best structure for sequential patterns modeling. For each block, AMEIR searches three operations:
\begin{align}
 \mathbf{H}_l = Act_l(Layer_l(Norm_l(\mathbf{H}_{l-1}))) + \mathbf{H}_{l-1},
\end{align}
where $\mathbf{H}_l$ is the concatenated sequential features from the $l^{th}$ layer. $Act_l, Layer_l, Norm_l$ stand for the selected operations for the $l^{th}$ layer, including activation function (\textit{e.g.}, \texttt{ReLU} or \texttt{GeLU}), layer operation (\textit{e.g.}, \texttt{CNN} or \texttt{RNN}), normalization (i.e., \texttt{Layer Normalization} or \texttt{Identical}). For non-sequential features, AMEIR adopts a similar progressive search strategy as AIM~\cite{aim} to search high-order feature interactions. %However, the interaction function of AMEIR is fixed as Hadamard Product. 
For MLP investigation, AMEIR searches the output dimension of each linear transformation layer. It defines a parameter matrix of the maximal output dimension as the search space and derives an optimal slice of each row as parameters of the corresponding layer of linear transformation. The widely-used one-shot weight-sharing random search paradigm is deployed to boost search efficiency in this work.

\noindent\textbf{Insight}.
1) The existing AutoML-based recommender systems evolve from single-component search to multi-component joint search, which facilitates reducing the introduction of manual experience and generating a unified model. 
2) Although frameworks introduced in this subsection search for multiple components of DRS comprehensively, most methods separately optimize each component. For example, AMEIR~\cite{ameir} searches three components in three steps with different strategies. This inefficient method may lead to sub-optimal DRS structures. DeepLight~\cite{deeplight} and OptInter~\cite{lyu2021memorize} provides a solution that we could integrate search spaces of different components to realize joint search;
3) Due to the high complexity of comprehensive search problems, researchers adopt various search strategies. Efficient gradient-based methods, including regularization methods, are mostly used~\cite{lyu2021memorize,deeplight,aim,umec}.

\section{Emerging Topics}
\label{emerging}
Section 4-8 present the application of the AutoML-based methods in various fields of the recommendation, such as input feature selection, feature embedding search, feature interaction search, and model training. Besides, automatic machine learning is also introduced in some emerging topics of recommendation system, such as GNNs-based recommendation, multi-modality recommendation, and \textit{etc}. In this section, we present these emerging topics.

\subsection{Automatic GNNs-based Recommendations}
Graph neural networks (GNNs) have been extensively studied recently due to their impressive capability in representation learning on graph-structured data~\cite{ding2021diffmg}. Recent works have been proposed to advance deep recommender systems based on GNNs techniques~\cite{fan2019graph,fan2020graph}. Thus, exploring the combination of AutoML and GNNs provides great opportunities to further boost the performance of GNNs-based recommendation methods.  
A few works have been proposed to study the combination of AutoML and GNNs together in the research community. 
For instance, GraphNAS~\cite{graphnas}  and Auto-GNN~\cite{autognn} made the very first attempt to enable the automatic design of the best graph neural architecture via reinforcement learning techniques. The main drawback of these two methods is computationally expensive because of the large search space.
To tackle this challenge, SANE~\cite{zhao2021search} proposes a differentiable architecture search algorithm for GNNs, where an advanced one-shot NAS paradigm is adopted to accelerate the search process.
As the first work to apply automatic NAS techniques to GNNs-based deep recommender systems, AutoGSR~\cite{chen2022autogsr} attempts to search for the optimal GNNs architecture on GNNs-based session recommendation through a differentiable architecture search algorithm.

\subsection{Automatic  Multi-Modality Recommendations}
%Personalized recommendations have become an important role in many online content sharing services, spanning from image, news, to video recommendations~\cite{wei2019mmgcn}. 
In addition to historical interactions between users and items, items' auxiliary knowledge from various modalities (\textit{e.g.}, visual, acoustic, and textual) has been incorporated to learn users' preferences for providing high-quality recommendation services~\cite{wei2019mmgcn}. 
Hence, it is desirable to advance deep multimodal learning via automated machine learning techniques~\cite{yin2021bmnas,perez2019mfas}, so as to design an optimal algorithm for targeted tasks. 
For example, as the first work on automated neural architecture search method for deep multimodal learning, the work of multimodal fusion architecture search (MFAS) aims to find accurate fusion architectures for multi-modal classification problem~\cite{perez2019mfas}.
%MMnas~\cite{yu2020deep} develops a generalized deep multimodal neural architecture search  framework consisting of a unified encoder-decoder backbone and task-specific heads for various multimodal learning tasks, where an efficient one-shot search algorithm is introduced to search the optimal architecture with weight sharing.
A bilevel multimodal neural architecture search framework (BM-NAS) is proposed to  learn the architectures of multimodal fusion models via a bilevel searching scheme~\cite{yin2021bmnas}. 

% 1)MFAS~\cite{perez2019mfas}
% \begin{itemize}
%     \item Search Space: Assume Modality Layers$\mathit{M,N}$, fusion layer number $\mathit{L}$,Operation $\mathit{P}$. Total Space size $(M\times N \times P)^L$.
%     \item Search Algorithm: RL(EPNAS,Designed by the same author, a combination of ENAS \& PNAS)
%     \item Evaluation: parameter sharing One-Shot
% \end{itemize}
% \noindent
% 2)BM-NAS~\cite{yin2021bmnas}
% \begin{itemize}
%     \item Search Space: Assume Modality Layers$\mathit{M,N}$. Upper Level $\mathit{C_{M+N}^{2}}$, Lower Level:?(Stop Condition is not given).
%     \item Search Algorithm: bi-level gradient-based DARTS with Sequential Setting for Both upper-level Search and Lower Lebel Search
%     \item Evaluation: parameter sharing One-Shot
% \end{itemize}
% \noindent
% 3)Deep Multimodal Neural Architecture Search~\cite{yu2020deep}
% \begin{itemize}
%     \item Search Space: Neural Architecture before fusion,Fusion Part is fitted.(Feature Extraction Part)
%     \item Search Algorithm: RL(ProxylessNAS)
%     \item Evaluation: parameter sharing One-Shot
% \end{itemize}
% \noindent
% 4)RamdomNet~\cite{alletto2020randomnet}
% \begin{itemize}
%     \item Search Space: Both Feature Extraction and Multi Modality Fusion
%     \item Search Algorithm: Random Search
%     \item Evaluation: parameter sharing One-Shot
% \end{itemize}
% \noindent

\subsection{Other Recommendation Tasks}
In addition to AutoML for GNNs-based and multimodal recommendations,  various important recommendation tasks are rarely explored through automated machine learning techniques, such as POI recommendations~\cite{zhao2020go}, sequential recommendations~\cite{kang2018self}, social recommendations~\cite{fan2019graph}, \textit{etc}. 
%For instance, next Point-of-Interest (POI) recommendations have been proposed to take advantage of the massive amounts of spatio-temporal (i.e., historical check-in) data in location-based social network (LBSN) services~\cite{zhao2020go},  such as Yelp and Uber. 
A few works have been conducted to apply automated neural architecture search techniques for spatio-temporal prediction such as AutoST~\cite{10.1145/3394486.3403122} and AutoSTG~\cite{10.1145/3442381.3449816}, which can help design  optimal neural architectures for POI recommendations. 
%In terms of sequential recommendations which aim to model users’ dynamic preferences from their historical behaviors~\cite{kang2018self}, the work of ~\cite{chen2021scene} introduces a sequential recommendation knowledge distillation (KD) framework  based on NAS. 
%with the development of online social networks services, social relationships among users are incorporated into enhancing users’ representations for recommendations~\cite{fan2019graph}. 
Besides, despite the success of various deep social recommendations, heavy manual work and domain knowledge is required to inherently combine user-item interactions and social relations~\cite{fan2020graph}, which can be addressed by AutoML techniques. % provide unprecedented opportunities to advance the design of deep social recommendations. 

\section{Future Directions}
Although many efforts have been made to design deep recommender systems from manual to automatic, there remain a few potential opportunities for future research directions. We first put forward some future directions from the perspective of summarizing the existing work in the above sections.

\begin{itemize}[leftmargin=*]
\item \textbf{Feature Selection.} 
Combinatorial features are of great importance for recommender systems due to the powerful capacity.
Therefore, how to generate and select combinatorial features effectively and save memory usage is an urgent problem for both industry and academics.
%The dropped feature fields of existing automated feature selection methods contain few of feature fields that cost a large amount of model parameters. As a result, they could enhance the model performance but hardly contribute to saving model parameters. How to conduct finer levels of selection, i.e, dropping a part of feature values rather than the whole feature field, is worth being explored.
\item \textbf{Feature Embedding Search.} As we all know, feature embeddings account for the majority of the parameters for the recommendation model. Therefore, combining the feature representation learning with model compression or quantization automatically may be a promising research direction.

\item \textbf{Feature Interaction Search.} Existing 
interaction functions, such as inner product and MLP, are widely-used in recommendation~\cite{guo2017deepfm,pnn}. Designing and introducing more informative interaction operators to generate more diverse interaction functions may improve the model prediction performance.

\item \textbf{Model Training.} The introduced AutoML-based works are mainly working on the training loss, including loss function selection or generation and regularization adjustment. More complex directions for model training are still unexplored such as optimizer settings, gradient direction guidance.
\item \textbf{Comprehensive Search.} Existing AutoML-based comprehensive search solutions search each component separately with heterogeneous search space~\cite{aim,autoias,ameir}, which results in low search efficiency and sub-optimal performance. Hence, it is potential to convert the heterogeneous search space into an isomorphic unified search space and perform efficient search algorithm.
\end{itemize}

Besides, we propose some further directions that are important to the recommender system but have not been explored yet.
\begin{itemize}[leftmargin=*]
\item As for the AutoML algorithm, existing work just directly transfers some algorithms~\cite{liu2018darts} widely used in CV or NLP to recommender systems.
However, simply applying these algorithms to personalized recommendation based on the sparse and high-dimensional user-item interaction data may be sub-optimal. 
Thus, it is important to design more suitable search and evaluation strategy for the recommendation scenario.

\item Multi-task learning~\cite{mmoe} that considers different revenue targets (such as click-through rate and conversion rate), is one of the most important techniques for industry recommendations. 
Thus, it is worthy of designing an automatic algorithm for adaptive multi-task learning.

\item User historical behaviors contain different dimensions of interests~\cite{dien}. Therefore, retrieving beneficial historical behaviors automatically to better model user preferences may be an important further direction.

\end{itemize}

\section{Conclusion}
Deep recommender systems have attracted increasing attention in both academia and industry. 
Besides, automated machine learning (AutoML), as one of the most promising AI techniques, has shown its great capabilities to advance deep architecture designs from manual to automatic. 
In this survey, we have conducted a comprehensive overview of an emerging research field: automated machine learning for deep recommender systems.  
Specifically, we first introduce the overview of AutoML techniques and deep recommender systems.
Then we discuss the state-of-the-art AutoML approaches that automate the feature selection, feature embeddings, feature interactions, and model training in DRS. %, and proposes some appealing research directions. 
We can observe from the development of these efforts, the AutoML-based recommender systems are developing to a \textbf{multi-component} joint search with \textbf{abstract} search space and \textbf{efficient} search algorithm.
We expect this survey can facilitate future research directions in the academic and industry community.

\bibliographystyle{ACM-Reference-Format}
\bibliography{0main}

% \appendix



\end{document}